\documentclass[aps,prx,twocolumn,amsmath,amssymb,nofootinbib]{revtex4-2}

\usepackage{graphicx,bm,amsmath}

\renewcommand{\figurename}{Figure}
\renewcommand{\tablename}{Table}
\makeatletter
\def\fnum@figure{\textbf{\figurename~\thefigure}}
\def\fnum@table{\textbf{\tablename~\thetable}}
\makeatother

\usepackage{etoolbox}
\patchcmd{\paragraph}{\itshape}{\bfseries}{}{}

\newcommand{\YRS}{YbRh\textsubscript{2}Si\textsubscript{2}}
\newcommand{\He}[1]{\textsuperscript{#1}He}
\renewcommand{\vec}[1]{\bm{\mathbf{#1}}}
\newcommand{\hvec}[1]{\hat{\vec{#1}}\vphantom{#1}}
\newcommand{\mtrx}[1]{\boldsymbol{#1}}
\renewcommand{\Re}{\operatorname{Re}}
\renewcommand{\Im}{\operatorname{Im}}
\DeclareMathOperator{\const}{const}

\begin{document} 
\let\oldaddcontentsline\addcontentsline
\renewcommand{\addcontentsline}[3]{}


\title{Odd-parity superconductivity underpinned by antiferromagnetism
in~heavy~fermion~metal~\YRS}

\author{Lev\ V.\ Levitin}
\email{l.v.levitin@rhul.ac.uk}
\affiliation{Department of Physics, Royal Holloway, University of London, Egham, UK}

\author{Jan Knapp}
\affiliation{Department of Physics, Royal Holloway, University of London, Egham, UK}

\author{Petra Knappov\'{a}}
\affiliation{Department of Physics, Royal Holloway, University of London, Egham, UK}

\author{Marijn Lucas}
\affiliation{Department of Physics, Royal Holloway, University of London, Egham, UK}

\author{J\'{a}n Ny\'{e}ki}
\affiliation{Department of Physics, Royal Holloway, University of London, Egham, UK}

\author{Petri Heikkinen}
\affiliation{Department of Physics, Royal Holloway, University of London, Egham, UK}

\author{Vladimir Antonov}
\affiliation{Department of Physics, Royal Holloway, University of London, Egham, UK}

\author{Andrew Casey}
\affiliation{Department of Physics, Royal Holloway, University of London, Egham, UK}

\author{Andrew\ F.\ Ho}
\affiliation{Department of Physics, Royal Holloway, University of London, Egham, UK}

\author{Piers Coleman}
\affiliation{Department of Physics, Royal Holloway, University of London, Egham, UK}
\affiliation{Department of Physics and Astronomy, Rutgers University, Piscataway, NJ, USA}

\author{Christoph Geibel}
\affiliation{Max Planck Institute for Chemical Physics of Solids, Dresden, Germany}

\author{Alexander Steppke}
\altaffiliation[Now at ]{Paul Scherrer Institute, Villigen, Switzerland}
\affiliation{Max Planck Institute for Chemical Physics of Solids, Dresden, Germany}

\author{Kristin Kliemt}
\affiliation{Physics Institute, Goethe University, Frankfurt, Germany}

\author{Cornelius Krellner}
\affiliation{Physics Institute, Goethe University, Frankfurt, Germany}

\author{Manuel Brando}
\affiliation{Max Planck Institute for Chemical Physics of Solids, Dresden, Germany}

\author{John Saunders}
\email{j.saunders@rhul.ac.uk}
\affiliation{Department of Physics, Royal Holloway, University of London, Egham, UK}


\date{24 January 2025}

\begin{abstract}
Topological superconductors are essential elements
of the periodic table of topological quantum matter.
However, the relevant odd-parity spin-triplet superconductors are rare.
We report high-resolution measurements of the complex electrical impedance of \YRS{}
down to 0.4\,mK, that reveal the presence of several superconducting states,
suppressed differently by magnetic field, both Pauli-limited and beyond the Pauli limit.
Superconductivity is abruptly switched off at the critical field of the primary antiferromagnetic order.
The onset of electro-nuclear spin density wave order enhances the superconductivity,
which we account for by the simultaneous formation of a spin-triplet pair density wave.
Together these observations provide compelling evidence for odd-parity superconductivity,
and its underpinning by antiferromagnetism, and allow us to identify the topological helical state.
\end{abstract}

\maketitle 

\section*{Introduction}
\vskip-0.5em

An unconventional superconductor is a macroscopic quantum state,
in which the wavefunction of the Cooper pairs varies over the Fermi surface,
breaking the symmetry of the underlying material~\cite{Norman2011}.
If the pair condensate order parameter has odd parity,
which in the simplest case of a spherical Fermi-surface corresponds
to pairs with orbital momentum $L = 1,3,\dots$,
then the system may support topological superconductivity of different classes,
in multiple superconducting states.
However, candidates for odd-parity superconductivity are rare.
Attention is currently focused on UTe\textsubscript{2}~\cite{Ran2019,Aoki2022} and
CeRh\textsubscript{2}As\textsubscript{2}~\cite{Khim2021}.
Other promising compounds include 
UGe\textsubscript{2}, UGeRh, UCoGe~\cite{Aoki2019review},
UBe\textsubscript{13}~\cite{Shimizu2019} and UPt\textsubscript{3}~\cite{Sauls1994,Gannon2015}.
UPt\textsubscript{3} is noteworthy for having three odd-parity superconducting phases,
one of which is a strong candidate for chiral superconductivity.

Odd-parity spin-triplet topological superfluid \textsuperscript{3}He exhibits both chiral and time-reversal-invariant phases with well-established order parameters~\cite{VWbook}.
The early identification of these superfluid states was possible because of: the intrinsic purity
and lack of disorder in this unique quantum fluid; the simple spherical Fermi surface
arising from the absence of crystal structure; weak spin-orbit coupling;
and the application of nuclear magnetic resonance to fingerprint the order parameters.
Magnetic field~\cite{VWbook} and anisotropic confinement in regular geometries~\cite{Levitin2013}
both alter the relative stability of these phases.
Moreover, new superfluid phases have been observed in \textsuperscript{3}He
in aerogels~\cite{Dmitriev2015,Halperin2019,Makinen2019,Dmitriev2021}.
These results illustrate the impact of symmetry-breaking fields
and disorder on odd-parity pairing states.

The present study focusses on the heavy fermion metal \YRS~\cite{Trovarelli2000}, with
tetragonal $D_{4h}$ symmetry~\cite{Krellner2012} and complex Fermi surface~\cite{Zwicknagl2016,Knebel2006}.
The electronic magnetism of \YRS{} is highly anisotropic,
reflected in a $g$-factor along the $c$ axis, which is an order of magnitude smaller
than that in the $ab$ plane~\cite{Sichelschmidt2003}.
The primary antiferromagnetic order AFM1, established at $T_N = 70$\,mK, has small
$\sim 0.002\,\mu_B$ staggered moments (here $\mu_B$ is the Bohr magneton)~\cite{Ishida2003}
and hitherto unresolved magnetic structure.
Superconductivity in \YRS{} was first revealed by the observation of the diamagnetic
screening, aligned with a magnetic transition at around 2\,mK~\cite{Schuberth2016}.
The onset of superconductivity at around 8\,mK has been subsequently
observed in electrical transport, but this study resolved no feature near 2\,mK~\cite{Nguyen2021}.
Recent heat capacity measurements reveal an electro-nuclear spin-density wave (SDW) order,
which we refer to as AFM2, below $T_A = 1.5$\,mK~\cite{Knapp2023},
stabilized by the strong hyperfine interactions of \textsuperscript{171}Yb and \textsuperscript{173}Yb.
Fig.~1A shows the magnetic phase diagram established for fields in the $ab$ plane~\cite{Knapp2025}.

\begin{figure*}[t]
\centerline{\includegraphics[scale=0.47]{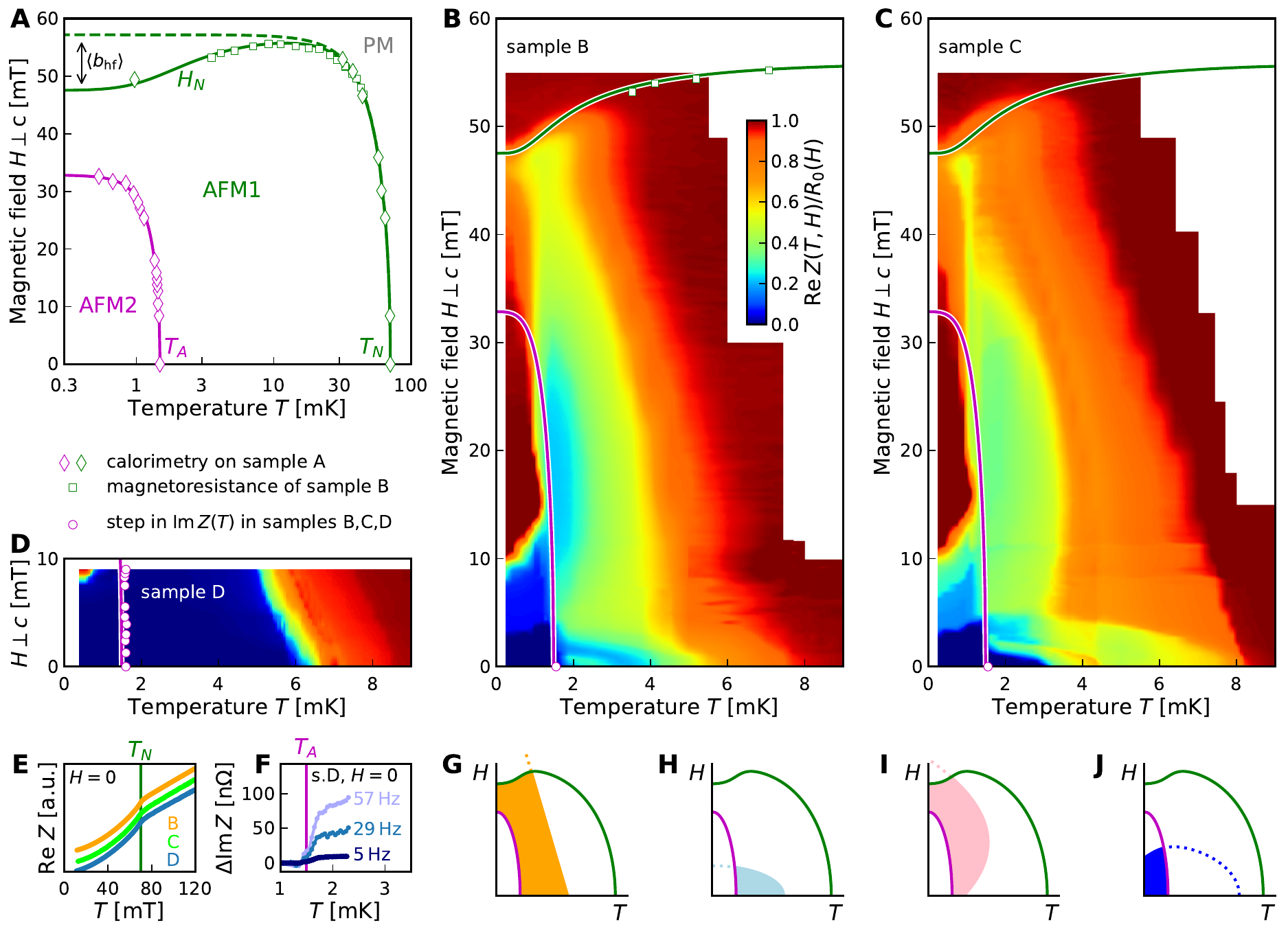}}
\caption{\textbf{Magnetic (A) and superconducting (B-D) phase diagram
of \YRS{} with in-plane magnetic field.}
\textbf{A}~Boundaries of the antiferromagnetic (AFM1 and AFM2) and paramagnetic (PM) phases
inferred from calorimetry and magnetoresistance~\cite{Knapp2023,Knapp2025}.
The back turn of the critical field $H_N$ of AFM1 below 15\,mK
is well accounted for by a hyperfine field
$\langle b_{\mathrm{hf}}\rangle$ exerted on Yb electrons by Yb nuclear spins, averaged over Yb isotopes.
Details in Ref.~\cite{Knapp2025} and SM.
\textbf{B}-\textbf{D}~The maps of the sample resistance $\Re Z(T,H)$ for 3 samples show
sample-to-sample variation, in contrast with the reproducible transport signature
of the N\'{e}el transition (\textbf{E}).
The AFM1/PM (green) and AFM1/AFM2 (magenta) phase boundaries are reproduced from \textbf{A}.
For sample~D the measurements are limited by the critical field of the Al contacts.
$\Re Z(T,H)$ is scaled by the normal state resistance $R_0(H) = \Re Z(11\,\mathrm{mK}, H)$.
We identify the observed sharp contours of constant resistance in \textbf{B}-\textbf{D}
with superconducting transitions in various parts of the heterogeneous samples.
The magnetic phase boundaries (solid lines reproduced from \textbf{A}) superimposed
onto \textbf{B}-\textbf{D} highlight abrupt suppression of superconductivity across
the AFM1/PM phase boundary and markedly different superconducting behavior inside
the AFM1 and AFM2 phases.
\textbf{F}~The superconducting signature of $T_A$ in the kinetic inductance $L_K = \Im Z(T) / \omega$, shown for sample~D at $H=0$. This feature as a function of field is marked in \textbf{B}-\textbf{D} with open magenta circles.
\textbf{G}-\textbf{J}~Classes of $\Re Z(T,H) = \const$ contours observed in the AFM1 and AFM2 phases across samples B,C,D and their potential extrapolation beyond the magnetic phase boundaries.}
\end{figure*}

Here we report the discovery of multiple superconducting states in \YRS,
revealed by high-resolution measurements of the complex electrical sample impedance
as a function of temperature and magnetic field (both in the basal $ab$ plane
and along the principal $c$ axis of the tetragonal structure, that we refer to as
``in-plane'' and ``out-of-plane'' respectively).
We establish that the superconductivity is always underpinned by antiferromagnetism in this system.
Weak nonuniformity in our single crystal samples
results in heterogeneous superconductivity, involving
distinct superconducting states, 
both Pauli-limited and exceeding the Pauli limiting magnetic field.
This unambiguously demonstrates spin-triplet pairing and allows us to identify
one of the pairing states as the topological helical phase. Remarkably this state
is abruptly boosted upon crossing from the electronic AFM1 phase into the electro-nuclear AFM2 phase.
We propose a mechanism for this boost mediated by a spin-triplet pair density wave (PDW).

While the extreme conditions of high magnetic field are important for
UTe\textsubscript{2}, superconductivity in \YRS{} occurs at ultra-low temperatures.
This study required the development of precise and low-dissipation SQUID-based
transport measurement techniques with n$\Omega$ resolution,
in-situ sample Johnson-Nyquist noise thermometry and a mode to observe flux quantization (see Methods).

\begin{figure*}[t!]
\centerline{\includegraphics[scale=1.13]{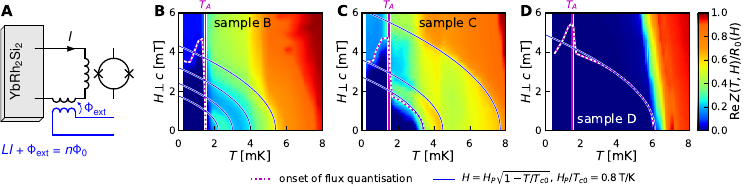}}
\caption{\textbf{Signatures of Pauli-limited superconductivity
in constant resistance and flux quantization contours.}
\textbf{A} The quantization of magnetic flux
in a loop comprising the \YRS{} sample and conventional superconductors
is detected by our SQUID-based transport measurement scheme. More in SM.
\textbf{B}-\textbf{D} Parabolic contours with a reproducible $H_P / T_{c0}$ ratio
are observed in all samples, despite strong sample-to-sample variations
and arbitrary field orientations in the $ab$ plane.
In samples B and C the onsets of zero resistance and flux quantization are aligned.
In contrast, sample~D exhibits zero resistance state beyond the Pauli limit.
Here the flux quantization is limited by the Pauli-limited suppression
of superconductivity in the contact regions.}
\end{figure*}

\section*{Results}

\paragraph*{Signatures of superconductivity with in-plane magnetic field $\vec H \boldsymbol{\perp c}$.}
We study the superconductivity in \YRS{} by measuring the complex electrical impedance $Z$ (see Methods).
Figure~1B-D shows the key result, the temperature-magnetic field maps of the resistance
(the real part of the impedance $\Re Z(T,H)$) of three single-crystal samples, B, C and D,
in relation to the magnetic phase diagram, Fig.~1A, determined from samples A and B.
The samples were selected for having similar residual-resistivity ratio $RRR \approx 50$,
and sharp reproducible signatures of the Ne\'{e}l transition
at $T_N = 70$\,mK~\cite{Gegenwart2002}, Fig.~1E.
The maps exhibit overall similarities: the onset of superconductivity around 8\,mK,
re-entrant normal state below $T_A$ at in-plane fields greater than about 10\,mT,
and abrupt suppression of the superconductivity at the critical field $H_N$ of the AFM1 phase.

Nevertheless, there are significant sample-to-sample variations,
that we understand in terms of heterogeneous superconductivity~\cite{Bachmann2019}:
$\Re Z$ drops below the residual normal-state value $R_0$ when superconducting regions appear
and $\Re Z = 0$ signifies the percolation of these regions.
Each sample exhibits distinct contours of constant resistance,
that we attribute to superconducting-normal phase boundaries in particular regions of the samples.
The diverse field dependencies signify different superconducting order parameters
stabilized in \YRS, and can discriminate between the many possible spin-triplet candidates.

A remarkable feature in $\Im Z(T)$ is the step-like drop at $T_A$, Fig.~1F,
coincident with the sharp heat capacity peak that manifests 
the bulk transition between AFM1 and AFM2 phases~\cite{Knapp2023}.
When fully imaginary, the impedance is proportional to the measurement frequency, Fig.~1F.
Thus, we attribute $\Im Z$ to the kinetic inductance $L_K \propto n_s^{-1}$~\cite{Meservey1969},
reflecting the superfluid density $n_s$ and geometry of the superconducting regions of the sample.
This drop at $T_A$ is a signature of a boost to superconductivity induced by the formation
of the electro-nuclear SDW of the AFM2 phase~\cite{Knapp2023}.
In samples C and D the $\Im Z(T)$ signature of $T_A$ is observed
on the background of zero resistance, Fig.~1C,D.
However, the resistance of sample~B only vanishes at $T_A$, Fig.~1B, indicating that,
in this case, complete percolation requires the onset of superconductivity in
previously normal regions of the sample, triggered by the boost at $T_A$.

\paragraph*{Multiple superconducting order parameters and in-plane $(\vec H \boldsymbol{\perp c})$ Pauli limit.}
The three classes of contours of constant $\Re Z(T,H)$ observed in the AFM1 phase are illustrated schematically in Fig.~1G-I:
linear (G), and parabolic (H) suppression of the critical temperature with field,
and non-monotonic field dependence (I). We focus on the second contour class, the parabolas,
that we associate with the Pauli (paramagnetic) limit~\cite{Clogston1962,Sarma1963} 
with a large Maki parameter $\alpha \gg 1$~\cite{Maki1966}.
Fig.~2 demonstrates a number of superconducting signatures with quadratic suppression
of the critical temperature $T_c(H) = T_{c0} (1 - H / H_P)^2$ with in-plane field $H$, all
with a reproducible ratio $\mu_0 H_P / T_{c0} = 0.8$\,T/K between
the critical temperature $T_{c0}$ at zero field and critical field $H_P$ at zero temperature.
In samples C and D similar parabolas describe another transport signature
inside the AFM1 phase: the onset of flux quantization, that demonstrates
the macroscopic phase coherence of the superconducting state(s), see Fig.~2A,C,D and Methods.

From the conventional expression for the Pauli limiting field
$\mu_0 H_P = \Delta \sqrt{2} / g\mu_B$ \cite{Clogston1962}
we obtain a reasonable value $\Delta = 0.7k_B T_{c0}$ of the energy gap, taking the in-plane
$g$-factor $g_{ab} = 3.5$ inferred from electron-spin resonance~\cite{Sichelschmidt2007}. 
In the presence of spin-orbit coupling, the pseudospin replaces the electron spin
as a good quantum number~\cite{Sauls1994,Mineev1999}.
This results in momentum-dependent $g$ factor of the quasiparticles involved
in pairing~\cite{Sauls1994}, which may affect the above gap estimate.
Corrections may also stem from the potential variations of the gap over the Fermi surface
and finite susceptibility in the $T = 0$ limit in case of (pseudo)spin-triplet pairing.

The two other contour classes we observe, Fig.~1G,I, previously reported in \YRS{} and
\textsuperscript{174}\YRS{}~\cite{Nguyen2021}, exhibit
significantly weaker suppression with in-plane field than the parabolas.
This demonstrates that the AFM1 phase of \YRS{} also hosts superconducting state(s)
exceeding the in-plane Pauli limiting field by an order of magnitude.

In contrast, in the AFM2 phase all samples exhibit just one class of
superconducting phase boundary, Fig.~1J, with critical field of order 10\,mT.
This field is comparable to, but larger than, the Pauli limit observed above $T_A$.
We propose that below $T_A$ only one odd-parity order parameter is stable,
the same as the state with in-plane Pauli limit above $T_A$, but with a boosted gap
and hence Pauli limiting field $H_P$.
The abrupt boost of $H_P$ at $T_A$ is particularly clear in the
$Re Z(T,H) = 0$ contour of sample~C and flux quantization contours of samples C and~D.

\paragraph*{Phase diagram with out-of-plane magnetic field $\vec H \boldsymbol{\parallel c}$\,.}
When the magnetic field is applied along the $c$ axis, Fig.~3,
the superconductivity extends up to 0.6\,T, close to
the estimated critical field $H_N$ of AFM1 for this field orientation (details in SM).
The abrupt switching off of the superconductivity near $H_N$ for both field orientations
together with the clear superconducting transport signatures of $T_A$ for $\vec H \perp c$,
Fig.~1B-D,F, unambiguously demonstrates that the superconductivity in \YRS{}
is tuned by the magnetic orders and is only stable in their presence.

In the absence of calorimetry data for $\vec H \parallel c$,
we map the clear superconducting signature of $T_A$ in $\Im Z(T)$, see Fig.~1F and SM.
This phase boundary extends up to $H_N$, in contrast with $\vec H \perp c$.
Moreover, below $T_A$ there are no signs of a re-entrant normal state,
exhibited by all samples with in-plane fields.

Taking the $g$ factor anisotropy
$g_{ab} / g_{c} = 20$~\cite{Sichelschmidt2007},
we estimate the Pauli limiting field along the $c$ axis to be as high as
0.1\,T above $T_A$ and 0.3\,T below $T_A$ (see SM). These values are below $H_N$,
suggesting that both in the AFM1 and AFM2 phases the
superconductivity exceeds the Pauli limit for $\vec H \parallel c$.
According to this analysis, below $T_A$ the single superconducting state
we identify has no Pauli limit for $\vec H \parallel c$;
above $T_A$ this property must be possessed by at least one of
the possible superconducting states revealed by the $\vec H \perp c$ results.

\begin{figure}[t]
\centerline{\includegraphics[scale=0.47]{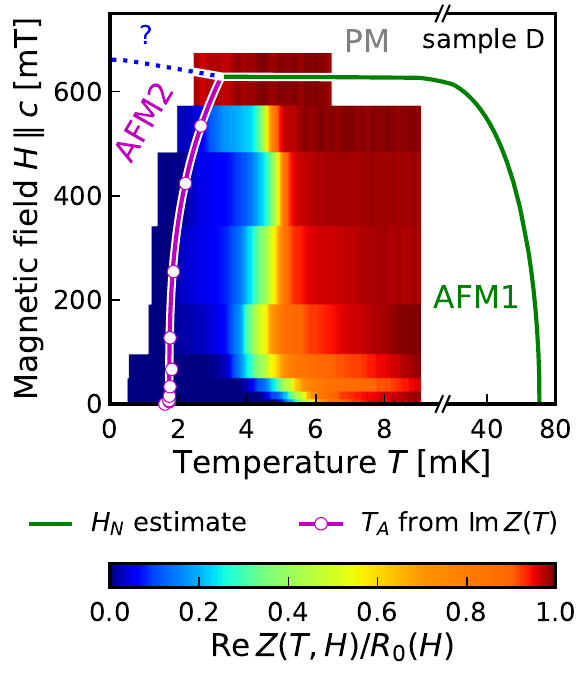}}
\caption{\textbf{Phase diagram of \YRS{} with out-of-plane magnetic field.}
The superconductivity is observed up to the fields close to the critical field $H_N$
of AFM1 (solid green line), estimated from our in-plane measurements (Fig.~1A) and the critical field anisotropy observed above 20\,mK~\cite{Gegenwart2002},
taking the hyperfine field into account (see SM).
The kinetic inductance signature of $T_A$ ($\Im Z(T)$) is observed all the way
up to $H_N$ and the re-entrant normal state below $T_A$ is absent,
in contrast to the in-plane fields, Fig.~1B-D.
Our data indicate the PM/AFM1/AFM2 polycritical point;
beyond it the PM/AFM2 phase boundary remains unchartered (shown schematically with a dotted blue line).}
\end{figure}

Despite strong anisotropy of magnetism and superconductivity in \YRS, the linearly-suppressed features, Fig.~1B,C,D and Fig.~3, exhibit similar initial slope $dT_c / dH$ for $\vec H \perp c$ and $\vec H \parallel c$.
Attributing this behavior to the conventional orbital suppression of superconductivity, we obtain a nearly-isotropic coherence length of approximately 100\,nm, consistent with a previous estimate~\cite{Nguyen2021}.
This large value is reasonable for a heavy fermion superconductor with the low $T_c$ observed in \YRS{}.

\section*{Discussion and Conclusions}

We now discuss the candidate superconducting order parameters in light of the observed
selective anisotropic Pauli limit and the boost in superconductivity at the onset
of electro-nuclear order spin density wave at $T_A$.
In a crystalline superconductor the possible order parameters are classified by the irreducible
representations (IRs) of the crystalline symmetry group~\cite{Volovik1985,Sigrist1991,Mineev1999}.
In the case of the $D_{4h}$ group, that describes \YRS{} above $T_N$~\cite{Krellner2012},
there are 5 odd- and 5 even-parity IRs.
Even-parity (pseudo)spin-singlet states would be Pauli-limited for all field orientations,
thus in view of the selective Pauli limit we
conclude that odd-parity (pseudo)spin-triplet pairing manifests in \YRS.
These states are described by the $\vec d(\vec k)$ vector,
the direction of zero spin projection at a given position $\vec k$ on the Fermi surface.
We now examine the 5 odd-parity IRs of the $D_{4h}$ group: one-dimensional $A_{1u}$, $B_{1u}$,
$A_{2u}$, and $B_{2u}$, and two-dimensional $E_u$~\cite{Volovik1985,Sigrist1991,Mineev1999}.
Strong spin-orbit coupling characteristic of heavy fermion metals
is predicted to further constrain $\vec d(\vec k)$ of each IR, either
to easy axis $\vec d \parallel c$ or easy plane $\vec d \perp c$~\cite{Sauls1994,Mukherjee2016}.
The $D_{4h}$ symmetry may be lowered by the antiferromagnetism, strain and magnetic field.
We assume the associated perturbation of the superconducting states we consider to be weak,
consistent with the absence of in-plane anisotropy of the Pauli limit, Fig.~2. 

We identify the superconducting states, that are Pauli limited for $\vec H \perp c$,
with the topological helical (planar) phase~\cite{Makhlin2014}.
Different orientations of the helical phase corresponding to the four one-dimensional IRs are 
$\vec d(\vec k) = \Delta(k_a\hvec a + k_b\hvec b)$ ($A_{1u}$),
$\vec d(\vec k) = \Delta(k_a\hvec b - k_b\hvec a)$ ($A_{2u}$)
$\vec d(\vec k) = \Delta(k_a\hvec a - k_b\hvec b)$ ($B_{1u}$),
and $\vec d(\vec k) = \Delta(k_a\hvec b + k_b\hvec a)$ ($B_{2u}$),
where the unit vectors $\hvec a$, $\hvec b$ and $\hvec c$ denote the principal crystallographic axes.
All four have isotropic Pauli limit in the $ab$ plane,
and no Pauli limit for $\vec H \parallel c$, consistent with the observations.
These states have identical gap structure, with point nodes at $\vec k \parallel c$
or nodeless on a cylindrical Fermi surface.
The $A_{2u}$ order parameter is distinguished 
by the predicted Josephson coupling to an s-wave superconductor along the $c$ axis~\cite{Jin2000}.
The observation of flux quantization requires a superconducting current between \YRS{} and
the Al contacts, Fig.~2A, and may point towards $A_{2u}$ (see SM).

\begin{figure}[t]
\centerline{
\includegraphics[scale=0.47]{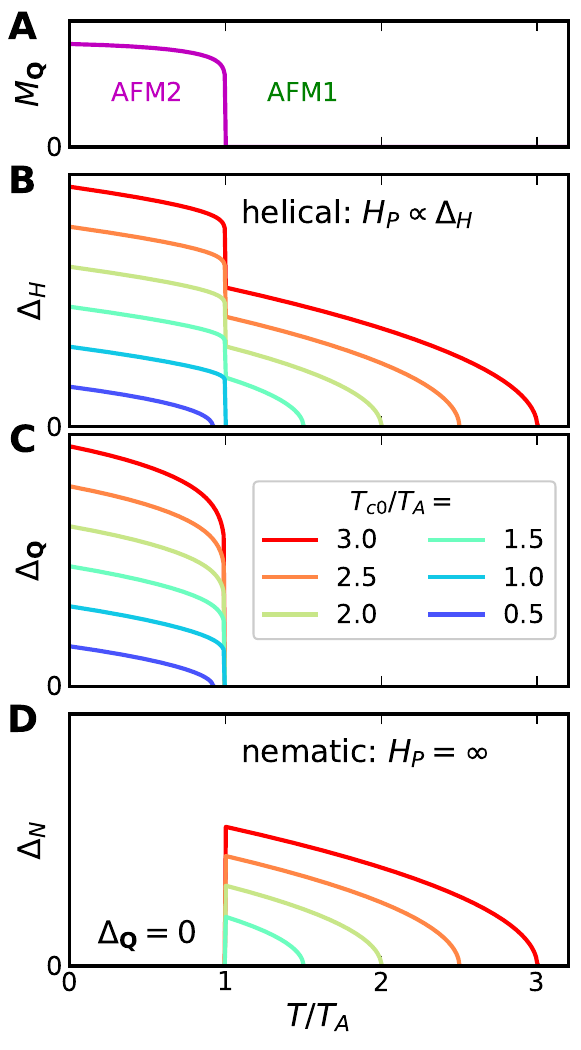}}
\vskip-0.5em
\caption{\textbf{Ginzburg-Landau model illustrating the boost of the helical superconducting order parameter upon entering the AFM2 phase}.
Below $T_A$ the magnetic order parameter $M_{\vec Q}$ of AFM2 (\textbf{A}) boosts the energy gap $\Delta_H$ of the helical superconducting state (\textbf{B}) via
the formation of a Cooper pair density wave with gap $\Delta_{\vec Q}$ (\textbf{C}). The Pauli limiting field, proportional to $\Delta_H$ is also boosted at $T_A$.
This attractive coupling is absent for the easy-plane nematic superconducting state, and competition between the superconductivity and AFM2 suppresses its gap $\Delta_N$ below $T_A$ (\textbf{D}).
The heterogeneity of the superconductivity is represented in this model by spatial variations of $T_{c0}$, the local transition temperature in the absence of $M_{\vec Q}$ order.
$M_{\vec Q}$ induces the helical superconducting state near $T_A$ in the regions of the sample where $T_{c0}$ is much smaller than $T_A$.}
\end{figure}

We further support the identification of the helical order parameter by
demonstrating how this superconducting state can be boosted at $T_A$, see Fig.~4 and SM.
At the heart of the boost mechanism is a spin-triplet PDW $\vec d_{\vec Q}(\vec k)$,
that couples to the helical order parameter $\vec d_H(\vec k)$ and
the staggered magnetization $\vec M_{\vec Q}$ of the AFM2 SDW via
$F = \lambda \langle i\vec d_H^* \times \vec d_{\vec Q}^{\vphantom{*}} + \text{H.c.}\rangle^{\phantom{|}}_{\vec k} \cdot \vec M_{\vec Q}$.
This interaction describes the diffraction of the Cooper pairs with amplitude $\vec d_H(\vec k)$
by the SDW, thereby forming a PDW $\vec d_{\vec Q}(\vec k) \propto \vec d_H(\vec k) \times \vec M_{\vec Q}$ of the same wavevector $\vec Q$, and lowering the condensate energy
by an amount $\propto \big\langle|\vec d_H(\vec k) \times \vec M_{\vec Q}|^2\big\rangle_{\vec k}$.
For instance with $\vec H \parallel a$
the $A_{2u}$ state acquires a 
modulation
$
\vec d(\vec k, \vec r) = \Delta_H \big(k_a\hvec b - k_b\hvec a\big) + i\Delta_{\vec Q} k_b \hvec c \cos(\vec Q \cdot \vec r).
$
In the regions of the sample with a pre-existing helical state,
its gap $\Delta_H$ and hence the Pauli limiting field
and superfluid density abruptly increase across $T_A$.
Moreover, in those regions where the tendency towards the helical order parameter is weak
($T_{c0} < T_A$), the superconductivity switches on at $T_A$ simultaneously with the PDW.
These two outcomes of the model successfully describe the superconducting transport signatures of $T_A$ observed in all samples, Fig.~1B,C,D.
This boost mechanism relies on the vector nature of the order parameters involved,
implying odd-parity spin-triplet pairing.

We now consider the regions of the sample where in the AFM1 phase
the superconductivity exceeds the in-plane Pauli limit.
If the easy-plane spin-orbit locking ($\vec d \perp c$), required by the helical state,
also applies here, we identify the nematic phase $\vec d(\vec k) = \Delta_N k_c \hvec d$ with arbitrary
in-plane orientation $\hvec d \perp c$. This state, characterized by a line node at $\vec k \perp c$,
belongs to the $E_u$ IR (details in SM). Like the analogous polar phase of the superfluid \He3~\cite{Dmitriev2015,Autti2016},
it may host half-quantum vortices.
To avoid the in-plane Pauli limit, $\hvec d$ must adjust to be perpendicular to the field.
The in-plane field is understood to have the same effect on $\vec M_{\vec Q}$~\cite{Knapp2023}.
Since both vectors are restricted to the $ab$ plane, $\vec M_{\vec Q} \parallel \hvec d$
and the nematic state receives no boost from the AFM2 SDW via the vector triple product interaction.
To account for the re-entrant normal state below $T_A$, our model includes
a direct competition between the superconductivity and SDW in the free energy (see SM),
that destabilizes the nematic state below $T_A$, Fig.~4D.

The superconductivity beyond the in-plane Pauli limit can also be accounted for
by any easy-axis ($\vec d \parallel c$) order parameter.
The key characteristic of this scenario is the out-of-plane Pauli limit.
However, for this field orientation the moderate Maki parameter $\alpha\sim 1$ 
prevents us from unambiguously identifying Pauli limited phase boundaries (more in SM).
A strong easy-axis candidate is the $E_u$ chiral phase
$\vec d(\vec k) = \Delta_C (k_a \pm i k_b)\hvec c$,
an analogue of the topological \He3-A~\cite{VWbook,Schnyder2015}.
Importantly, the chiral and helical states have the same gap structure $|\vec d(\vec k)|$
and are found to be nearly-degenerate on a quasi-2D Fermi surface~\cite{Ng2000,Annett2006}.
Such competition between chiral and helical states has been discussed in the context of
Sr\textsubscript{2}RuO\textsubscript{4}, a tetragonal unconventional superconductor
previously considered to be spin-triplet~\cite{Chronister2021}.

Two experimental observations, unusual for superconductors, remain beyond our model.
In the AFM1 phase the enhancement of $T_c$ with magnetic field, Fig.~1B,C,I and Fig.~3,
qualitatively resembles the phase diagram of UTe\textsubscript{2}~\cite{Aoki2022}
and invites explanation in terms of field-dependent strength of the pairing interactions~\cite{Nguyen2021,Mineev2020};
at high field we cannot rule out superconducting order parameters
different from the states we identify at low fields.
In the AFM2 phase the decrease of the Pauli limiting field by several mT
on cooling below $T_A$, Fig.~1B,C,D,J, may be connected to the hyperfine effects,
that are also responsible for the back turn of $H_N(T)$, Fig.~1A~\cite{Knapp2025}.


A major outstanding question is: what makes the superconductivity in \YRS{} heterogeneous?
Candidates include crystalline defects, with density varying on a scale of the coherence length, local strain and magnetic domains.

The abrupt destruction of the superconductivity in \YRS{} at the critical field
of AFM1, discussed as a quantum critical point, calls into question
the proposed significance of quantum criticality
for the pairing mechanism~\cite{Schuberth2016,Nguyen2021,Schuberth2022}.
The interplay of the superconductivity with antiferromagnetic orders
rather suggests spin-fluctuation-mediated pairing~\cite{Monthoux2007},
in which the balance between ferromagnetic and antiferromagnetic
spin fluctuations plays an important role.
Both types of spin fluctuations are well established
in \YRS{}~\cite{Gegenwart2002,Ishida2002,Sichelschmidt2003,Stock2012,Wolfle2009}.

The triplet PDW driven by the AFM2 SDW, proposed in this work, is related to
the singlet and triplet PDWs stabilized by pre-existing charge density waves (CDW),
discussed in NbSe\textsubscript{2}~\cite{Liu2021} and UTe\textsubscript{2}~\cite{Gu2023} respectively.
A defining characteristic of \YRS{} is that we have access to superconductivity
both with and without the SDW order. 
We stress that, although $T_A$ is below the onset of superconductivity,
the hyperfine energy scale of the AFM2 SDW is much higher than the superconducting pairing energy,
so the superconductivity is simply responding to this new magnetic order (see SM).

In conclusion, we demonstrate odd-parity spin-triplet superconductivity and its strong interplay with two AFM orders in \YRS{}.
We observe multiple superconducting states 
and identify one of them with the topological helical phase, with a high degree of confidence.
We show how a spin-triplet PDW can mediate the boost to the superconductivity at the onset of the electro-nuclear SDW.

The challenge for \YRS{} is to stabilize uniform bulk superconductivity, e.g.\ 
by improved sample quality, elimination of residual strain~\cite{Bachmann2019},
application of uniaxial strain~\cite{Steppke2022,Panja2024}.
Future goals include phase-sensitive measurements to confirm and complete
the identification of triplet order parameters,
and the investigation of the emergent surface states in the candidate
topological superconducting phases using scanning tunneling microscopy~\cite{Gu2025}.

In the wider context of heavy fermion superconductivity, the rarity of superconductivity
in Yb-based heavy fermion compounds, 4f hole analogues of Ce-based 4f electron heavy
fermion metals~\cite{Weng2016}, and the low critical temperatures of the two discovered examples,
$\beta$-YbAlB\textsubscript{4}~\cite{Nakatsuji2008} and \YRS{}, remains a mystery.
Nevertheless, the low transition temperatures in \YRS{} confer advantages through modest energy scales:
the magnetism and superconductivity are effectively tuneable by magnetic field, uniaxial strain
and Yb isotopic substitution. The relatively long coherence length makes \YRS{} suitable for
single-crystal superconducting quantum devices with nanobridge junctions.
In the light of the increased accessibility of ultra-low temperature platforms~\cite{Nyeki2022}
and associated techniques there is future promise in this flexibility, coupled
with the new technological opportunities offered by topological superconductivity. 



%
%

%
%

\section*{Acknowledgments}
We acknowledge invaluable discussions with Seamus Davis, James Sauls and Philip Brydon,
and thank Paul Bamford, Richard Elsom, Ian Higgs and Harpal Sandhu
for excellent technical support.

\textbf{Funding.}
This work was supported by the European Microkelvin Platform, EU’s H2020 project under grant agreement no. 824109, 
Deutsche Forschungsgemeinschaft (German Research Foundation) through grants Nos.\ BR 4110/1-1, KR3831/4-1, TRR 288 (422213477, project A03),
the US National Science Foundation grant DMR-1830707 (PC)
and Leverhulme Trust Research Project Grant (AFH).
We thank the Wilhelm and Else Heraeus Foundation for support through a Heraeus Visiting Professor position at Goethe University Frankfurt.

\textbf{Authors contributions.}
LL, JS and MB supervised the project;
KK and CK grew the single crystals of \YRS{};
LL, JK, MB, AS, VA, AC, ML and JN designed and prepared the ultra-low temperature measurement setups;
LL, JK, PK, ML, PH and  JN performed the experiments;
LL analyzed the data with input from JS, JK, CG and MB;
PC and AFH developed the Ginzburg-Landau theory of superconductivity coupled to SDW and PDW;
LL examined the candidate triplet order parameters with input from PC, AFH, JK and JS;
all authors discussed the results and interpretation;
LL and JS wrote the manuscript with input from all authors.

\textbf{Competing interests.} The authors declare that they have no conflict of interests.

\textbf{Data and materials availability.} All data supporting this research are available in the Main Text, Supplementary Materials and at \texttt{https://figshare.com/s/093b2afe3e12c7ca72b6}.


\section*{Supplementary materials}
\noindent Materials and Methods\\
Supplementary Text\\
Figs.~S1 to S15\\
Table~S1\\
References \cite{Kliemt2020,Knapp2024ultcal,Drung2007,Casey2014,Levitin2022,Heikkinen2024,Hamann2019,Werthamer1966,Maki1964,Yip1993,Surovtsev2019,LL:IX}


\clearpage

\pagebreak


\newcommand{\mSDW}{M}
\newcommand{\unitmSDW}{m}
\newcommand{\mSC}{S}

\newcommand{\Zs}{Z}
\newcommand{\alignedZs}{Z_{\phantom{s}}}
\newcommand{\Rs}{R}

\newcommand{\FS}{_{\vec k}^{\,}}

\newcommand{\um}{\textmu{}m}


\renewcommand{\thefigure}{S\arabic{figure}}
\renewcommand{\thetable}{S\arabic{table}}
\renewcommand{\theequation}{S\arabic{equation}}
\setcounter{figure}{0}
\setcounter{table}{0}
\setcounter{equation}{0}

\onecolumngrid
\pagebreak

\begin{center}
\Large Odd-parity superconductivity underpinned by antiferromagnetism\\
in heavy fermion metal \YRS. Supplementary Materials
\smallskip

\large \centerline{Lev V.\ Levitin, Jan Knapp, Petra Knappov\'{a}, Marijn Lucas, J\'{a}n Ny\'{e}ki, Petri Heikkinen,}
\centerline{Vladimir Antonov, Andrew Casey, Andrew Ho, Piers Coleman, Christoph Geibel,}
\centerline{Alexander Steppke, Kristin Kliemt, Cornelius Krellner, Manuel Brando, John Saunders}

\end{center}
\tableofcontents

\pagebreak

\twocolumngrid

\let\addcontentsline\oldaddcontentsline


%

\pagebreak
\onecolumngrid

\begin{figure*}[t!]
\includegraphics[width=0.8\textwidth]{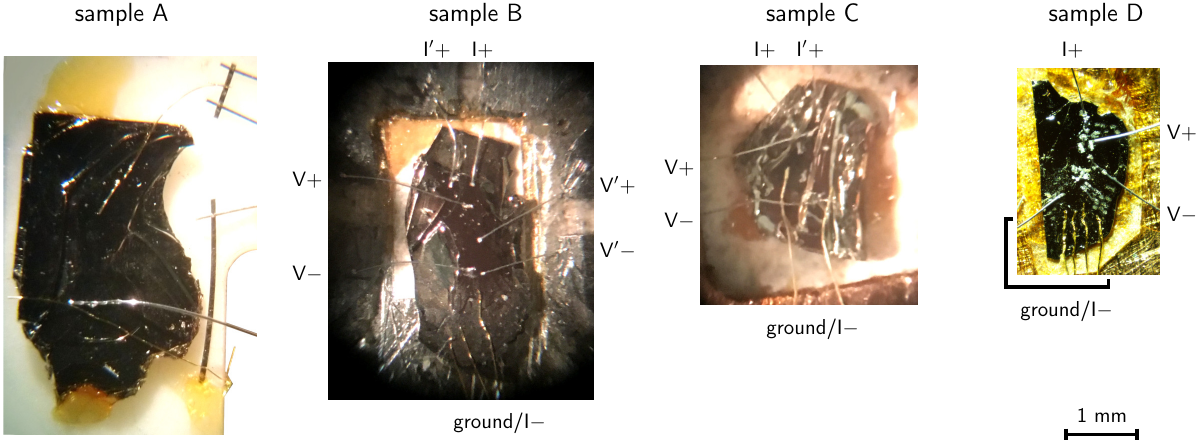}
\caption{Photographs of the four samples used in this work. Sample A was studied in a calorimeter. See Ref.~\cite{Knapp2024ultcal}
for the details of the technique. Samples B-D are shown configured for electrical transport measurements with current source I$+$,
voltage probes V$\pm$ and thermal ground/current sink I$-$.
Samples B and C were equipped with spare current I$'+$ and voltage V$'\pm$ contacts.}\label{fig:samples}
\end{figure*}

\twocolumngrid

\section{Materials and Methods}

\subsection{Single crystal samples}

This work is based on studies of four single crystals of \YRS{} with residual resistivity ratio $RRR \approx 50$~\cite{Krellner2012,
Kliemt2020}, see Fig.~\ref{fig:samples}.
Sample A was probed with calorimetry~\cite{Knapp2023,Knapp2024ultcal,Knapp2025}.

Samples B-D were configured for 4-terminal electrical transport measurements, described in the next section.
These crystals were attached to copper (samples B,D) and silver (sample C) sample holders using GE varnish via a layer of cigarette paper.
Each of these samples was thermally attached to its holder using 2-4 25\,\um{} Au wires, spot-welded to one side of the sample
and ultrasonically bonded to the holder, also serving as a current sink I$-$.
The current source I$+$ and voltage V$\pm$ probes were connected by ultrasonic bonding
25\,\um{} Al wires to the sample and to Nb pads on the sample holder.
In the case of sample D a similar Al wire shunted the gold wires (thermal link), intended as a superconducting current sink below the critical field of Al.
After the initial measurements with in-plane fields, this sample was re-mounted for studies in
out-of-plane fields. We then made new Al contacts, leading to a slight change in the observed properties.

\subsection{Electrical impedance measurements}

\begin{figure*}[t!]
\centerline{\includegraphics[width=0.8\textwidth]{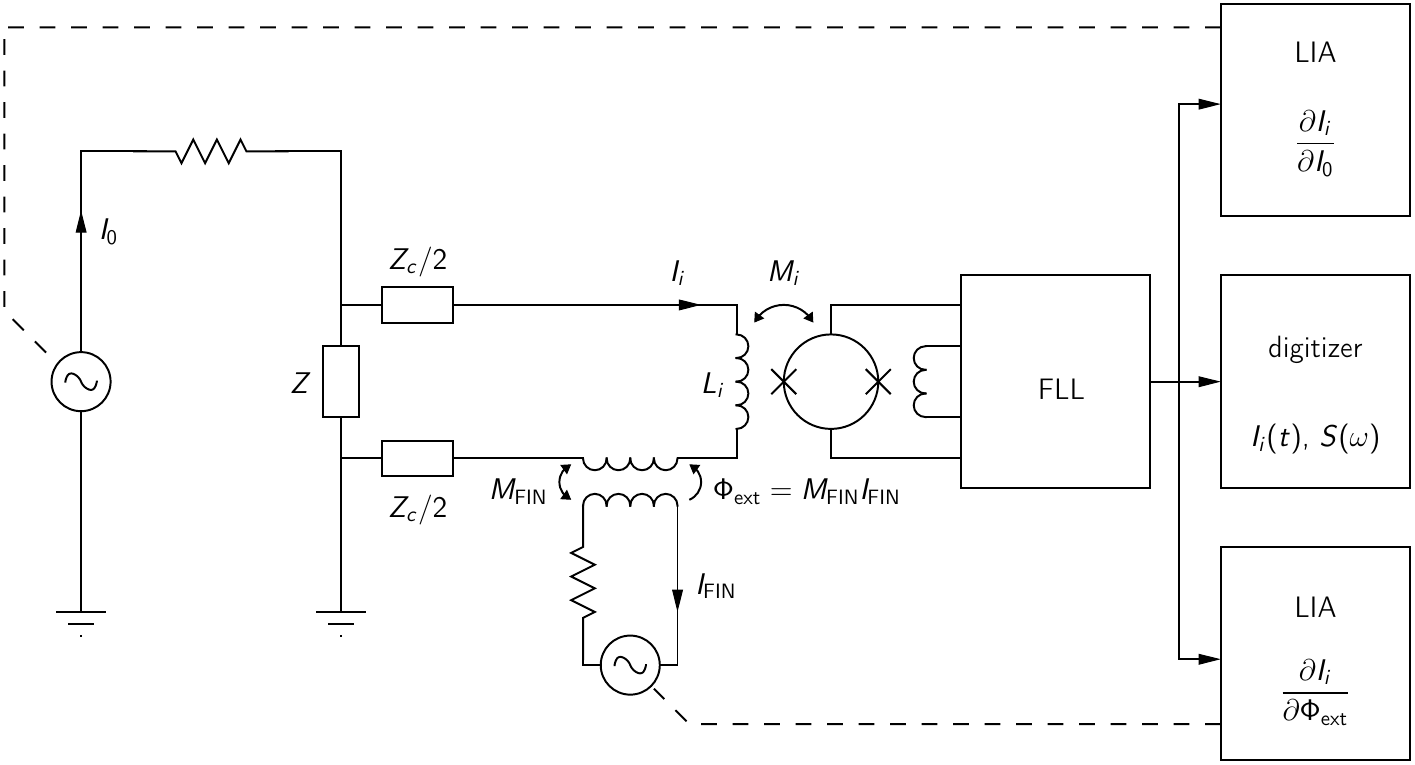}}
\caption{SQUID impedance meter. The sample $\Zs$ is connected to a dc SQUID current sensor via contacts $Z_c$.
Current $I_i$ in the SQUID input coil $L_i$ is amplified using flux-locked loop (FLL) electronics.
The circuit is driven with current $I_0$ and flux $\Phi_{\text{ext}}$.
The responses measured with two lock-in amplifiers (LIA) yield $\Zs$ and $Z_c$
via Eqs.~\eqref{eq:Zs} and \eqref{eq:Zc}.
A digitizer measures $I_i(t)$ in time domain
and the power spectral density $S(\omega)$ of the $I_i$ noise.
These are used respectively to observe the magnetic flux quantization, Fig.~\ref{fig:fluxq},
and measure sample temperature (when $\Re(\Zs + Z_c) > 10\,\mu\Omega$).}\label{fig:setup}
\end{figure*}

We have developed a technique for low-dissipation measurements of the complex sample impe\-dance $Z$
and contact impedance $Z_c$.
The latter includes contributions from the $V+$ and $V-$ Al wires as well as the regions of the sample adjacent to the voltage probes.
We connect the voltage probes to a SQUID current sensor, Fig.~\ref{fig:setup},
and measure the phase-resolved response of the input current $I_i$ of the SQUID
to two harmonic drives $I_0$ and $\Phi_{\text{ext}}$.
We either apply the two drives sequentially and obtain $\partial I_i/\partial I_0$ and $\partial I_i/\partial \Phi_{\text{ext}}$ at the same frequency $\omega$,
or make simultaneous measurements at two nearby frequencies $\omega$ and $\omega'$.

\begin{figure*}[t!]
\centerline{\includegraphics[scale=0.47]{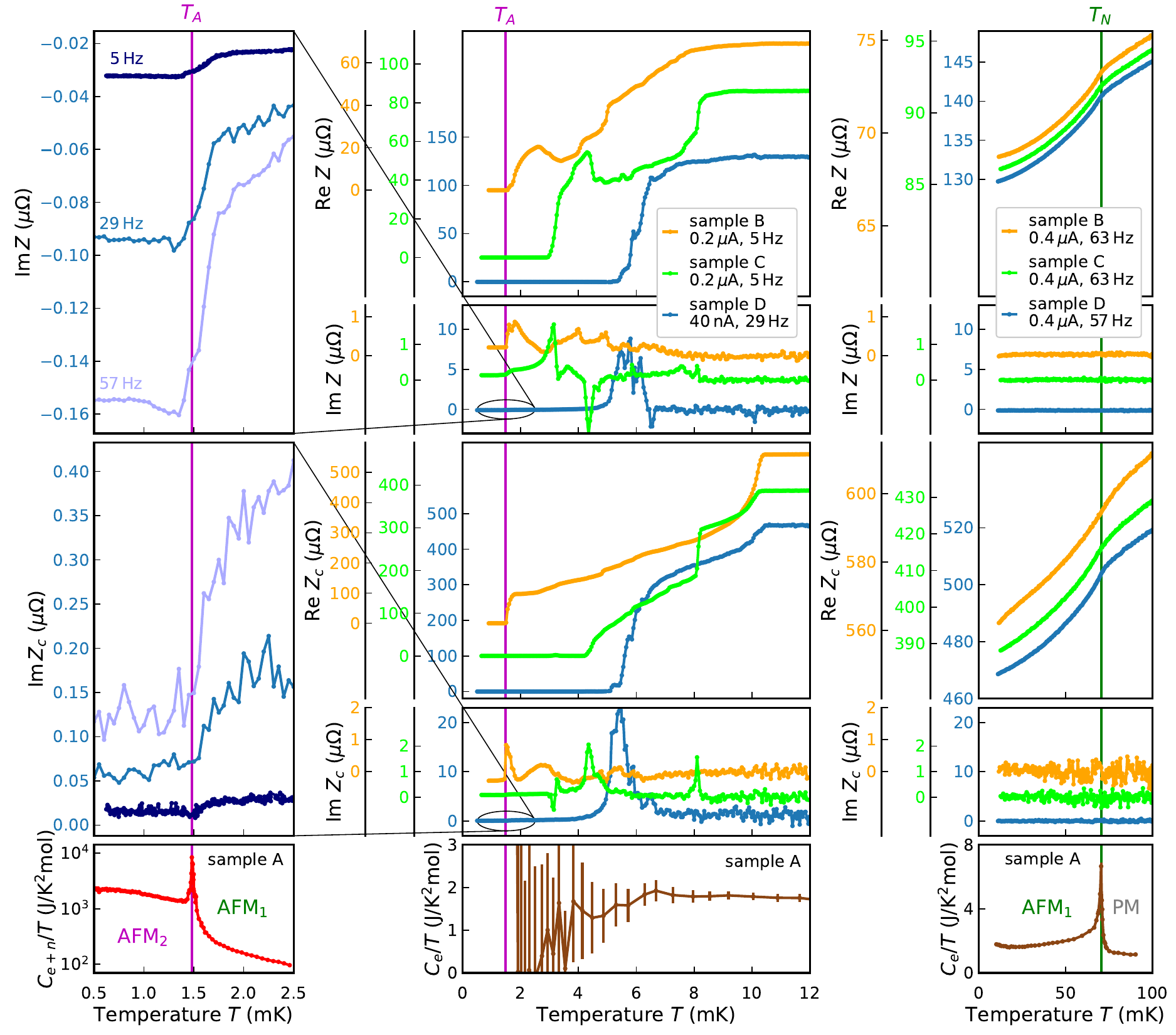}}
\caption{\label{fig:zerofield}Transport signatures of superconductivity and antiferromagnetism in samples B, C and D and their contacts, in comparison with heat capacity signatures in sample A.
Here all measurements are at zero (the Earth's) field.}
\end{figure*}

The response to flux $\Phi_{\text{ext}}$ injected via a superconducting transformer determines the total impedance of the SQUID input loop
\begin{equation}\label{eq:Zt}
\Zs + Z_c + i\omega L_i = -i\omega \left/ \frac{\partial I_i}{\partial\Phi_{\text{ext}}}\right.,
\end{equation}
where $L_i \approx 1\,\mu$H is the inductance of the input coil of the SQUID.
The drive current $I_0$ is divided between $\Zs$ and $Z_c + i\omega L_i$, $I_i$ being the current in the latter branch,
and from Kirchhoff's laws
\begin{equation}\label{eq:Zs:general}
\Zs = \frac{\partial I_i}{\partial I_0}\big(\Zs + Z_c + i\omega L_i\big).
\end{equation}
Thus we obtain the impedances
\begin{eqnarray}
\alignedZs &\!\!=\!\!& -i\omega\frac{\partial I_i}{\partial I_0} \left/ \frac{\partial I_i}{\partial\Phi_{\text{ext}}}\right.,\label{eq:Zs}\\
Z_c &\!\!=\!\!& -i\omega\left(1 - \frac{\partial I_i}{\partial I_0} \right) \left/ \frac{\partial I_i}{\partial\Phi_{\text{ext}}}\right. - i\omega L_i.\label{eq:Zc}
\end{eqnarray}
Figure~\ref{fig:zerofield} illustrates the transport measurements in Earth's magnetic field.

Unlike conventional 4-terminal resistance measurements, in our scheme a non-zero current $I_i$ is drawn via the voltage probes.
Nevertheless, $\Zs$ obtained from Eq.~\eqref{eq:Zs} agrees with
the value which would be obtained by the conventional method,
with a high-impedance voltmeter connected to the sample instead of the SQUID current sensor.
To show this we note that, for any given current drive $I_0$, simultaneously applying a flux drive
\begin{equation}\label{eq:Zs:voltmeter}
\Phi_{\text{ext}} = -I_0 \frac{\partial I_i}{\partial I_0} \left/ \frac{\partial I_i}{\partial\Phi_{\text{ext}}}\right.
\end{equation}
nulls the current through the voltage probes and the SQUID input coil, $I_i = 0$. Then the current
$I_s = I_0$ through the sample and voltage $V_s = i\omega \Phi_{\text{ext}}$ across it yield the conventional sample impedance $\Zs = V_s/I_s$, coincident with the value given by Eq.~\eqref{eq:Zs}.

The above circuit analysis requires $\Zs$ and $Z_c$ to be current-independent.
We find this to be the case, except close to abrupt jumps in $\Re \Zs(T)$ and $\Re Z_c(T)$,
were the critical current in some regions of the sample/contacts is comparable to $I_0$.
To mitigate the errors in $\Zs$ and $Z_c$ associated with the nonlinearity, we simultaneously drive $\Phi_0$
at a nearby frequency $\omega'$ while measuring $\partial I_i / \partial I_0$, and then swap the frequencies to measure $\partial I_i / \partial\Phi_{\text{ext}}$. Thus the current distribution in the sample and contacts remains approximately unchanged.



\subsubsection{Johnson-Nyquist noise measurements and sample temperature}

When the sample and contacts chain is resistive, the circuit acts as a current-sensing noise thermometer (CSNT) \cite{Casey2014}.
In the simplest case when both $\Zs$ and $Z_c$ are predominantly real and frequency-independent,
the Johnson-Nyquist voltage fluctuations in the total resistance $R_t = \Zs + Z_c$ result
in noise current $I_i$ with power spectral density
\begin{equation}\label{eq:CSNT}
S(\omega) = \frac{4 k_B R_t T}{R_t^2 + \omega^2 L_i^2} + S_0(\omega).
\end{equation}
Here $k_B$ is Boltzmann's constant, $T$ is the average temperature of the sample and contacts
weighted by their resistances, and $S_0(\omega)$ is the background noise.
By fitting $S(\omega)$ to Eq.~\eqref{eq:CSNT} we obtain $T$ and $R_t$ simultaneously.
This technique can be applied simultaneously with lock-in transport measurements
by discarding frequencies close the drive frequency from the $S(\omega)$ fit.

The temperature of the sample is measured directly, when the contacting Al wires
(and the rest of the connections to the SQUID input coil, formed from Nb and NbTi) are superconducting.
Practically the noise measurements are limited to frequencies above $\omega_{\min} / 2\pi \approx 1$\,Hz,
because of parasitic low-frequency noise.
As a result the fits to to Eq.~\eqref{eq:CSNT} can only reliably yield $T$ and $R_t$,
when $R_t > L_i \omega_{\min} \sim 10\,\mu\Omega$.

\begin{figure}[b!]
\centerline{\includegraphics[scale=0.47]{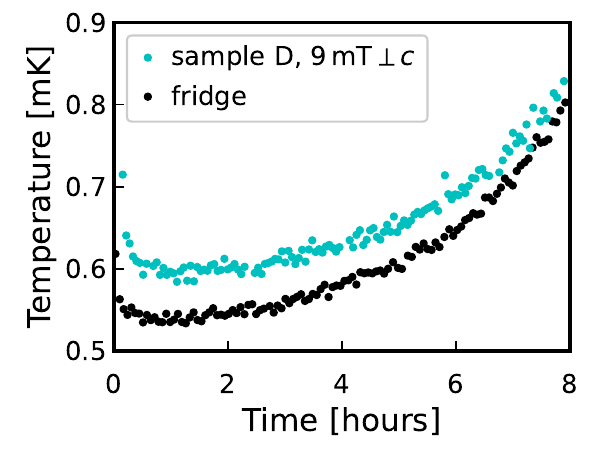}}
\caption{\label{fig:Tyrs}Current-sensing noise thermometry
on sample D in re-entrant normal state, simultaneous with a lock-in transport
measurement with $I_0 = 40$\,nA drive at 27\,Hz.
Good thermalization is observed down into the microkelvin regime.}
\end{figure}

\begin{figure}[t!]
\centerline{\includegraphics[scale=0.47]{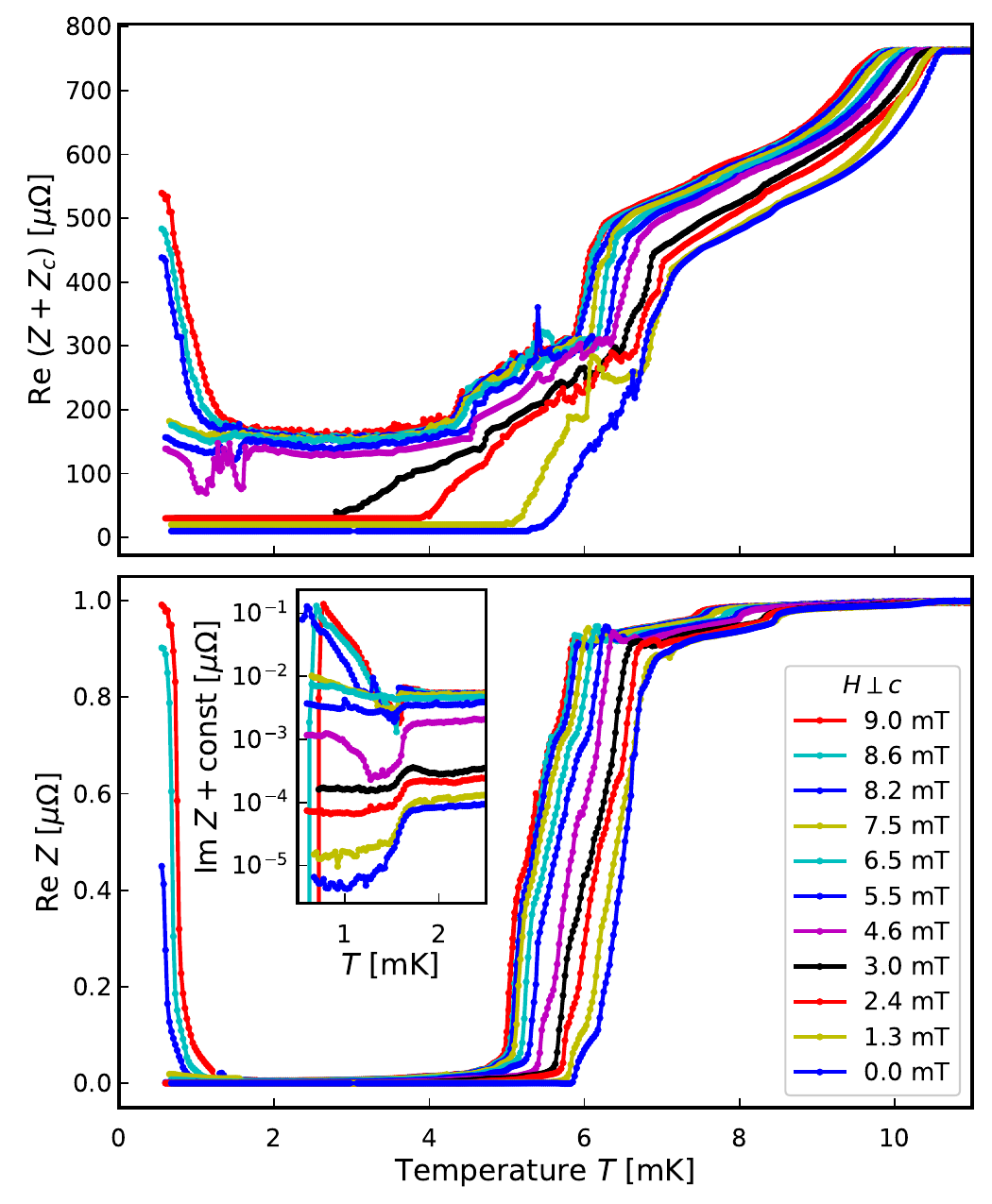}}
\caption{\label{fig:sampleD:Hab}Combined driven and noise measurements on sample D with in-plane magnetic fields.
Due to the lack of $\Phi_{\text{ext}}$ drive, the impedance was obtained using Eqs.~\eqref{eq:ZsRt}
and \eqref{eq:RsRcLi}.}
\end{figure}

Sample D enters the re-entrant normal state, see Figs.~1D, \ref{fig:sampleD:Hab}
below the critical field of Al contacts.
This enabled the direct sample noise thermometry well below 1\,mK, Fig.~\ref{fig:Tyrs}.
We demonstrate that the temperature gradient between the sample and nuclear demagnetization
refrigerator platform is at most 0.1\,mK, corresponding to sub pW heat leak to the sample,
even when running a driven transport measurement.

Sample noise thermometry is not possible in the important regime
of zero sample and contact resistance, in particular near $T_A$.
Since we argue that the superconductivity in our YbRh$_2$Si$_2$ samples is heterogeneous,
it is expected to have no significant effect on the thermal conductance of the samples.
Thus we routinely infer the sample temperature (except for sample D above 100\,mT,
see Sect.~\ref{sect:sampleD:Hc} below)
from the precise noise and $^3$He melting curve thermometers
installed on the refrigerator platform, with a systematic uncertainty of $\sim 0.1$\,mK.
In addition to any overheating of the sample with respect to the refrigerator due to the heat load on it,
this figure includes the temporal lag between the sample and refrigerator,
in particular near $T_A$, where the thermalization time constant exceeds 1 hour.

We argue that this procedure demonstrated in sample D, Fig.~\ref{fig:Tyrs},
is also valid for samples B and C, prepared and thermalized in the same way.
The key difference is the addition of the $\Phi_{\text{ext}}$ drive, which
produces a current in the sample comparable to $I_0$.
Both drive lines are interrupted with Ag epoxy filters
to reduce heating due to noise in the room-temperature electronics~\cite{Levitin2022}.

Above the critical field of Al wires (5-10\,mT depending on the orientation of the wires with respect to the field due to demagnetizing effects) the contact impedance exhibits complicated field, temperature and frequency dependence, potentially related to superconductivity in a material formed out of some subset of Al, Yb, Rh and Si near the Al-YbRh$_2$Si$_2$ interface.
Eq.~\eqref{eq:CSNT} does not apply from the onset of superconductivity in Al up to fields of about 30\,mT.
At higher fields the contacts are purely resistive and the use of noise thermometry is recovered.

\begin{figure}[t!]
\centerline{\includegraphics[scale=0.47]{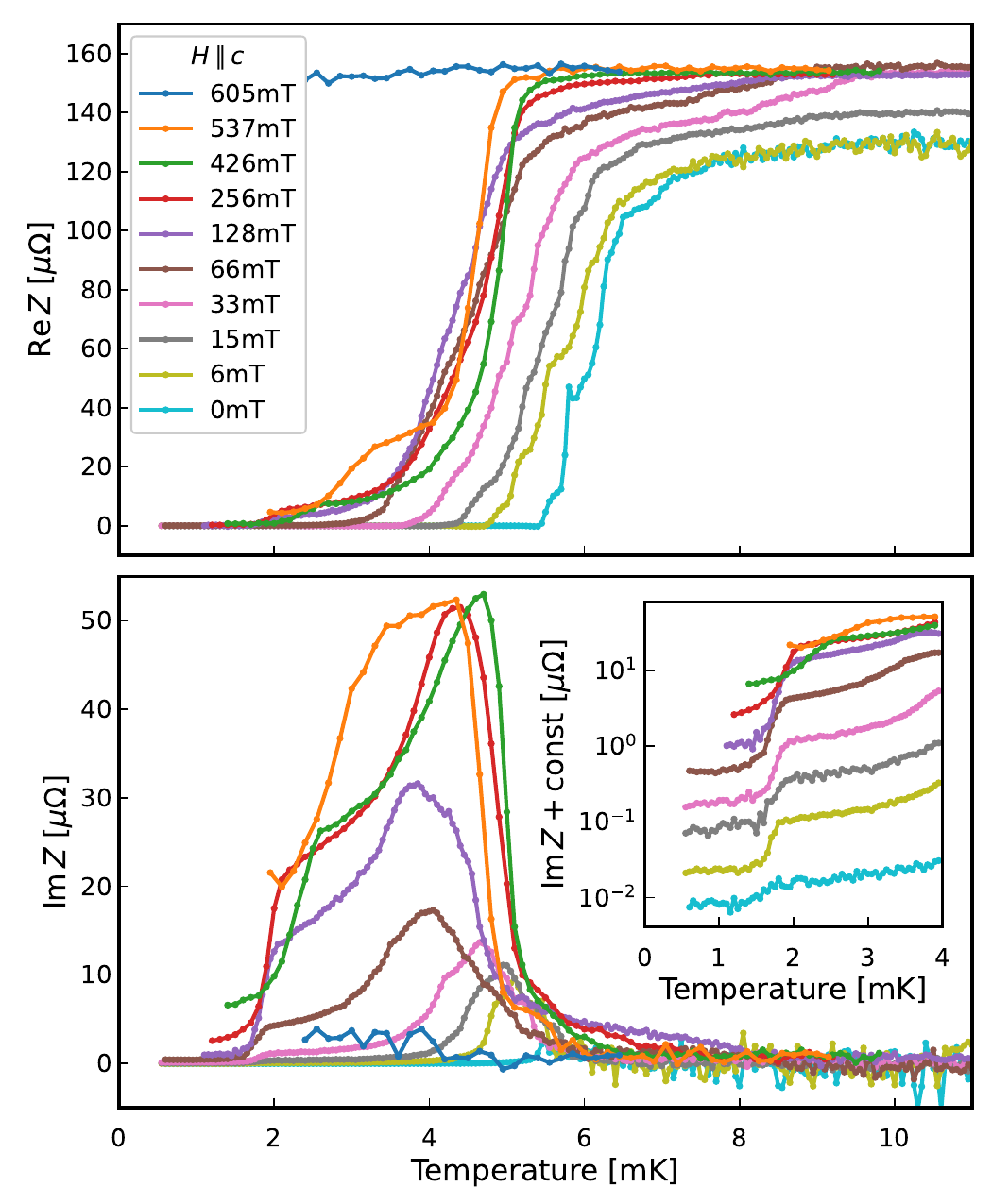}}
\caption{\label{fig:sampleD:Hc}Transport measurements on sample D with out-of-plane magnetic fields.
The measurements below 100\,mT were obtained in the same setup as the rest of the transport study and
the sample temperature is assumed to be equal to that of the refrigerator.
A different setup was used above 100\,mT with a significant temperature gradient between the sample
and refrigerator plate. Here the temperature is inferred from sample noise thermometry, Eq.~\eqref{eq:CSNT}, with the temperature dominated by the Al wirebonds.
The inset illustrates the transport signatures of $T_A$ in $\Im Z(T)$ at various magnetic fields.}
\end{figure}

\begin{figure*}[t!]
\vskip-1em
\centerline{\includegraphics[scale=0.47]{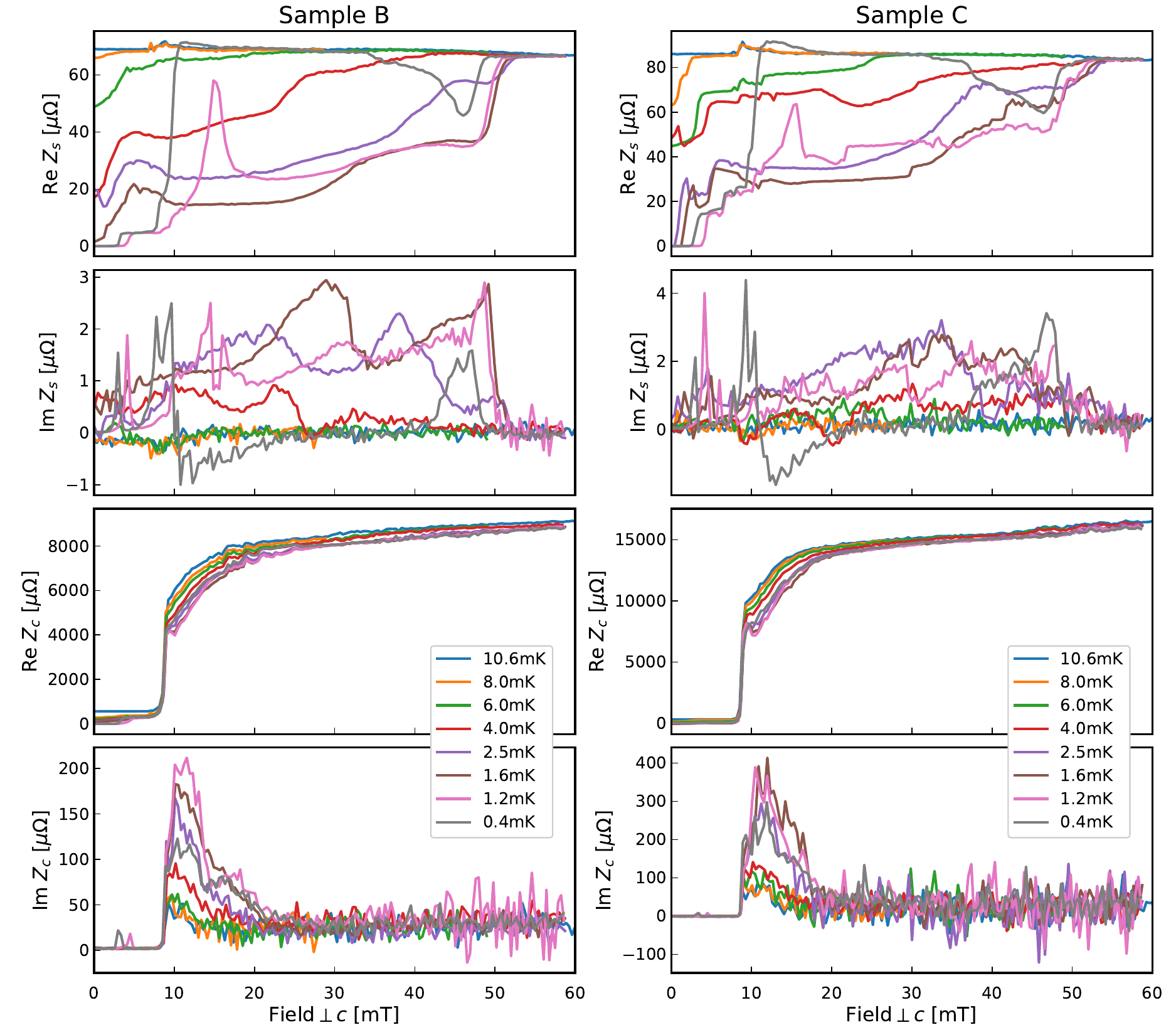}}
\caption{\label{fig:samplesBC}Transport measurements on samples B and C with out-of-plane magnetic fields,
performed during step-wise field sweeps with current drive $I_0 = 200$\,nA at 5\,Hz.
The contact impedance abruptly changes at the critical field of Al connections
(with a demagnetizing factor $\sim 0.8$), which has a weak systematic effect on the sample impedance.
}
\end{figure*}

\subsubsection{Sample D. Measurements with in-plane fields}

The impedance measurement scheme presented above was used for the measurements on samples B and C,
and on sample D under magnetic field applied along the $c$~axis.
In the historical sequence of measurements sample D was first studied with fields $< 10$\,mT applied in the $ab$~plane,
and here no $\Phi_{\text{ext}}$ drive was implemented. However,
the noise spectra were routinely captured simultaneously with the driven measurements,
and from these we inferred using Eq.~\eqref{eq:CSNT} a detailed map of $R_t(T,H)$, see Fig.~\ref{fig:sampleD:Hab}.
To improve the resolution in $R_t$ and remove the $R_t > 10\,\mu\Omega$ limit
here we constrained $T_{\text{sample}} = T_{\text{fridge}}$.

We observe that in other samples $\omega L_i \gg \Im (\Zs + Z_c)$ at low magnetic fields and approximate
$\Zs + Z_c + i\omega L_i \approx R_t + i\omega L_i$ in Eq.~\eqref{eq:Zs:general} to obtain
\begin{equation}\label{eq:ZsRt}
\Zs = \big(R_t + i\omega L_i\big) \frac{\partial I_i}{\partial I_0},\qquad
Z_c = R_t - \Re \Zs
\end{equation}
from a combination of driven and noise measurements.

When $\Im \Zs \ll \Re \Zs$ we switch to a different method, that assumes $\Zs$ and $Z_c$ to be purely resistive. Under this assumption the sample and contact resistances $\Rs$ and $R_c$ can be obtained
from the driven measurements alone, improving the resolution over Eq.~\eqref{eq:ZsRt}.
To do so we rewrite Eq.~\eqref{eq:Zs:general} as
\begin{equation}\label{eq:RsRcLiKirch}
\frac{R_c}{\Rs} + \frac{i\omega L_i}{\Rs} = 1 \left/ \frac{\partial I_i}{\partial I_0}\right. - 1,
\end{equation}
and obtain
\begin{equation}\label{eq:RsRcLi}
\Rs = \frac{\omega L_i}{\displaystyle\Im \left(1 \left/ \frac{\partial I_i}{\partial I_0}\right.\right)},\qquad
R_c = \Rs \Re \left(1 \left/ \frac{\partial I_i}{\partial I_0}\right. - 1\right).
\end{equation}
The temperature sweeps shown in Fig.~\ref{fig:sampleD:Hab} were obtained on warming.
Although vortex pinning is likely to be significant in a heterogeneous superconductor,
we observe no difference between field-cooled and zero-field-cooled measurements.
This is consistent with Ref.~\cite{Schuberth2016}, were such a difference in
magnetic measurements was only prominent at fields below 0.1\,mT, an order of magnitude smaller
than the typical fields that affect the sample impedance $Z$.


\subsubsection{Sample D. Measurements with out-of-plane fields}\label{sect:sampleD:Hc}

To complete the study of sample D with out-of-plane fields it was necessary to apply fields in excess of 100\,mT, see Figs.~2, \ref{fig:sampleD:Hc}.
Such fields are beyond the capability of the small home-made sample magnets,
used for all other measurements. Instead we utilised a large commercial NMR magnet on
a different nuclear demagnetization refrigerator~\cite{Heikkinen2024}.
In order to minimise the disruption to ongoing NMR experiments,
the sample holder was relatively weakly connected to the distant nuclear demagnetization stage
with no local thermometer capable of operating above 30\,mT.
Therefore we employed the sample noise thermometry. At these relatively high fields the contacts were purely resistive and dominated $R_t$. This is a potential source of error, since we measure the contact rather than sample temperature.
The key conclusions of Fig.\ 2 are insensitive to systematic errors in the temperature scale
at fields above 100\,mT.

Importantly Fig.~\ref{fig:sampleD:Hc} shows the signature of $T_A$ in $\Im Z$ over the entire field range.
These $T_A$ are plotted in Fig.~3.
Also shown in Fig.~\ref{fig:sampleD:Hc}, the normal state sample resistance gradually increases from
6 to 30\,mT. We attribute this to the change in the current distribution between Al and Au I- wires,
see Fig.~\ref{fig:samples}, as the Al wire and Al-YbRh$_2$Si$_2$ contacts are driven normal by the field.
Similar parasitic effect was not observed in samples B and C, Fig.~\ref{fig:samplesBC},
which only had Au I- wires.

\subsubsection{Samples B and C. Magnetic field sweeps}

The $Z(T,H)$ maps for samples B and C were obtained from high-resolution field sweeps
at constant fridge temperatures. The data were taken at fixed fields, with the sample magnet in persistent mode. At each field we measured the response to $I_0$ drive at 5\,Hz and $\Phi_{\text{ext}}$ at 7\,Hz first, then swapped the frequencies, then changed the field by 0.1-0.4\,mT.
Between the field sweeps the temperature was changed by 0.1-0.3\,mK.
Example sweeps are shown in Fig.~\ref{fig:samplesBC}

\subsection{Observation of flux quantization}\label{sec:fluxq}

\begin{figure}[h!]
\centerline{\raisebox{15em}{\hskip1em\includegraphics[width=10em]{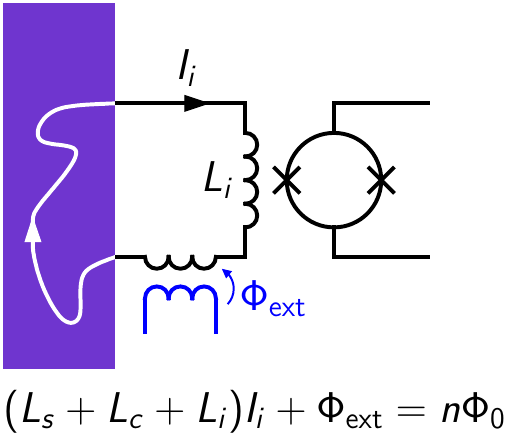}}~~~%
\hskip-11em\includegraphics[scale=0.47]{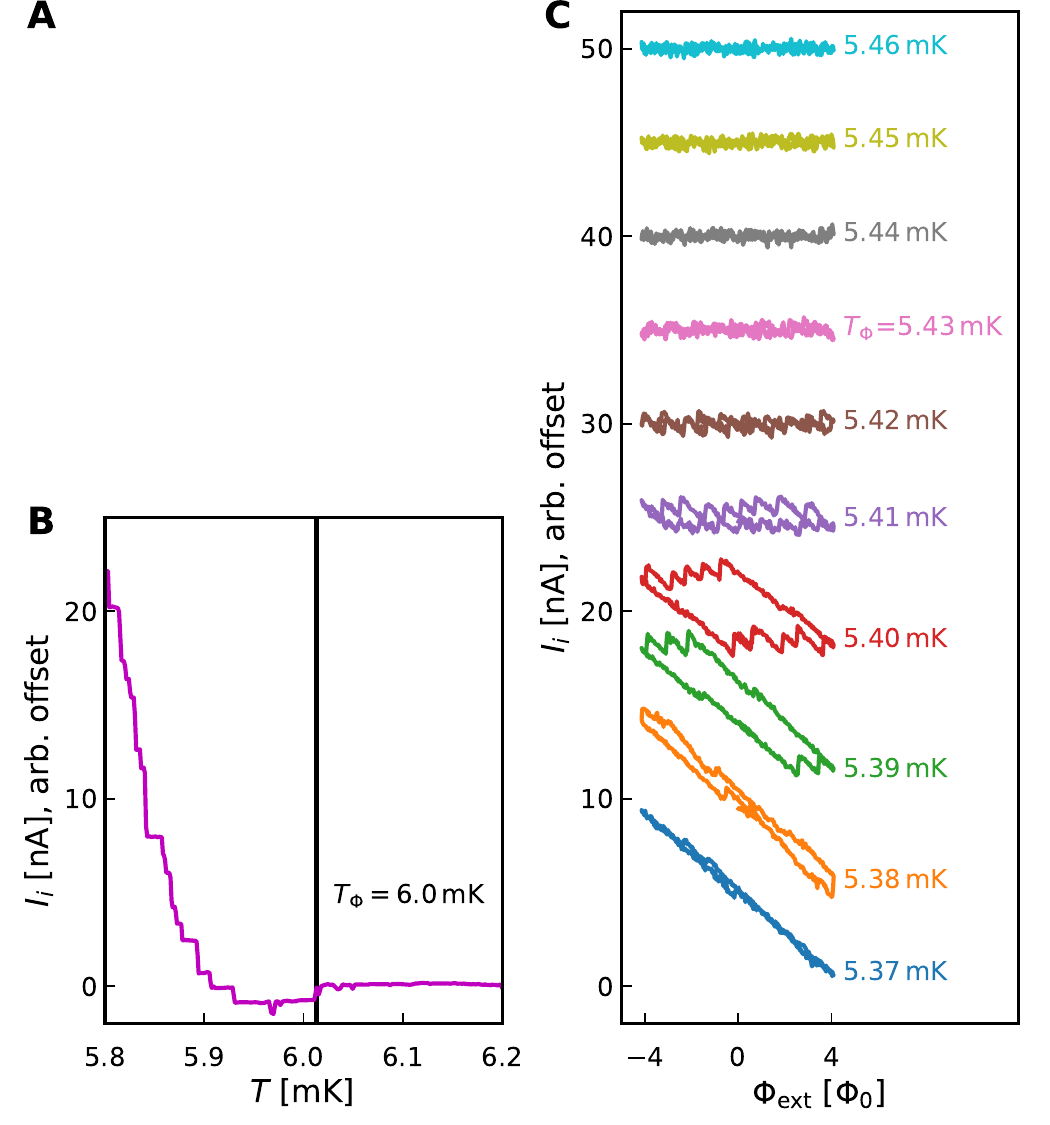}}
\caption{Observation of flux quantization. \textbf{A}
Flux quantization is observed in a loop formed by the sample, the input coil of the SQUID (Nb)
and the secondary coil of the $\Phi_{\text{ext}}$ transformer (Nb),
complete with Al, Nb and NbTi connections.
The static magnetic field $\vec H$ applied to the sample contributes an unimportant constant offset to $\Phi_{\text{ext}}$.
\textbf{B} Quantized jumps in SQUID input current $I_i$ measured on a slow warm-up in zero field
(Earth's magnetic field). The size of the jumps corresponds to the flux in the loop
changing by $(1.0 \pm 0.1)\Phi_0$.
The flux is induced at low-temperature by the residual magnetic field. 
This technique was used in the studies of sample D under in-plane magnetic field,
where the $\Phi_{\text{ext}}$ drive was not implemented.
\textbf{C} When the superconducting loop is complete, $\Phi_{\text{ext}}$ modulation ($8\Phi_0$ peak-to-peak triangular wave at 50\,mHz) induces a reproducible change in $I_i$, as observed at 5.37\,mK.
The critical current of the loop reduces on warming, leading to jumps in $I_i$ similar to those observed in \textbf{B}. Shown here are the measurements on sample D during a slow warm-up in zero (Earth's) magnetic field during the study with out-of-plane fields. The onset temperature $T_{\Phi}$
of flux quantization is different in \textbf{B} and \textbf{C} due to re-mounting
and re-contacting the sample between these experiments.}\label{fig:fluxq}
\vskip7em
\end{figure}

The two methods of observation of flux quantization are illustrated in Fig.~\ref{fig:fluxq}.

\section{Phase diagram}

\subsection{Antiferromagnetic phase diagram}

\begin{figure}[b!]
\includegraphics[scale=0.47]{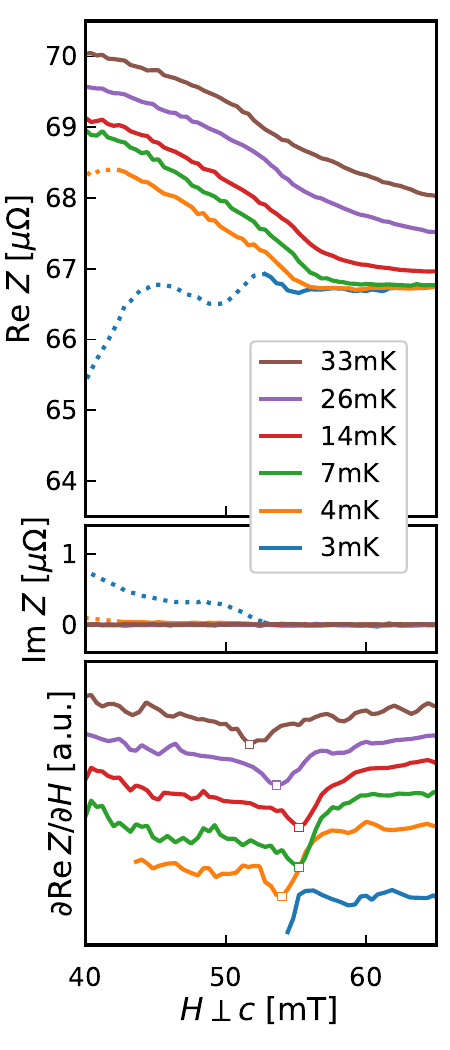}\hfill%
\parbox[b]{0.25\textwidth}{\caption{Magnetoresistance signatures of the critical field $H_N$ of AFM1.
In the normal state \YRS{} exhibits negative magnetoresistance.
Between 4 and 45\,mK our data reveal a clear minimum in $\partial \Re Z/\partial H$.
We identify this feature with $H_N$, since above 30\,mK it coincides with the heat capacity
signature of the N\'{e}el transition, see Fig.~\ref{fig:AFM1}.
We use the $\Im Z < 0.05\,\mu\Omega$ condition (data shown with solid lines)
as the criterion of the sample being fully normal.
When this condition is not met (dotted lines), the field dependence of $\Re Z$ is dominated
by the suppression of superconductivity near $H_N$.
This prevents the observation of the magnetoresistance signature of $H_N$ below 4\,mK.}\label{fig:MR}}
\end{figure}

\begin{figure*}[t!]
\centerline{\includegraphics[scale=0.47]{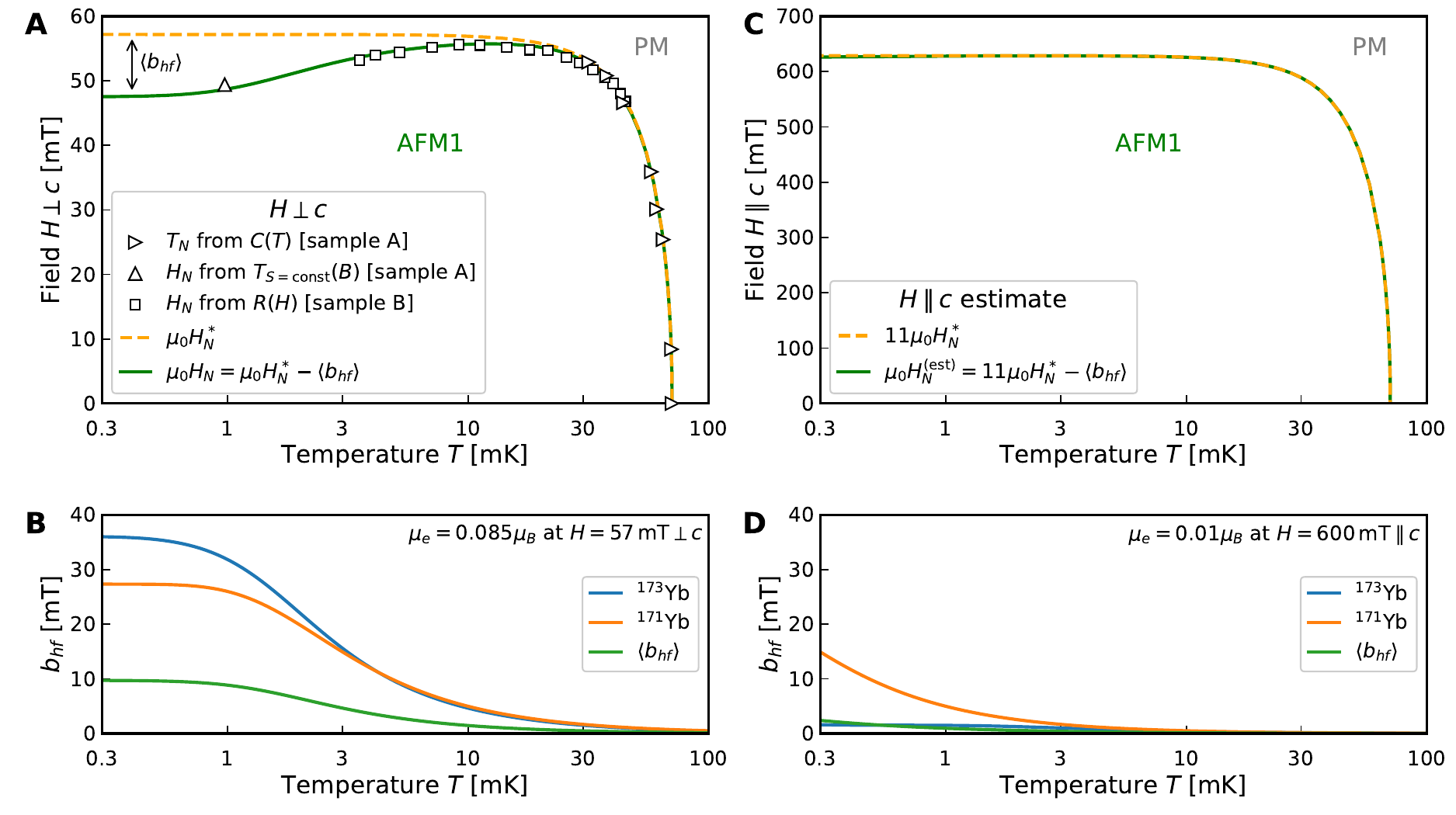}}
\caption{AFM1/PM phase boundary for two principal field orientations.}\label{fig:AFM1}
\end{figure*}

For in-plane fields the AFM1/PM and AFM1/AFM2 phase boundaries $H_N(T)$ and $T_A(H)$,
shown in Figs.~1 and \ref{fig:SC:contours}, are based on sharp peaks in high-resolution heat capacity measurements~\cite{Knapp2023}.
In both cases we use other techniques to extend the data to lower temperatures~\cite{Knapp2025}.

\subsubsection{In-plane AFM1/AFM2 phase boundary}

In the case of $T_A(H)$ this is done using the magnetocaloric signatures observed in the same experiment down to 0.5\,mK~\cite{Knapp2024ultcal,Knapp2025},
shown in Fig.~1A with diamonds, similar to the heat capacity peaks.
The AFM1/AFM2 boundary plotted in Figs.~1 and \ref{fig:SC:contours} is based on a fit to these combined data
\begin{align}\label{eq:TA:fit}
\big(\mu_0 H_A(T)\big)^2 = &-204.409\,\mathrm{mT^2/mK^4} \times T^4 \notag\\
& - 34.981,\mathrm{mT^2/mK^2} \times T^2 \notag\\
& + 1080.046\,\mathrm{mT^2},
\end{align}
where $H_A(T)$ is the inverse function of $T_A(H)$.
\vskip8em

\subsubsection{In-plane AFM1/PM phase boundary}

\begin{figure*}[p!]
\centerline{\includegraphics[scale=0.47]{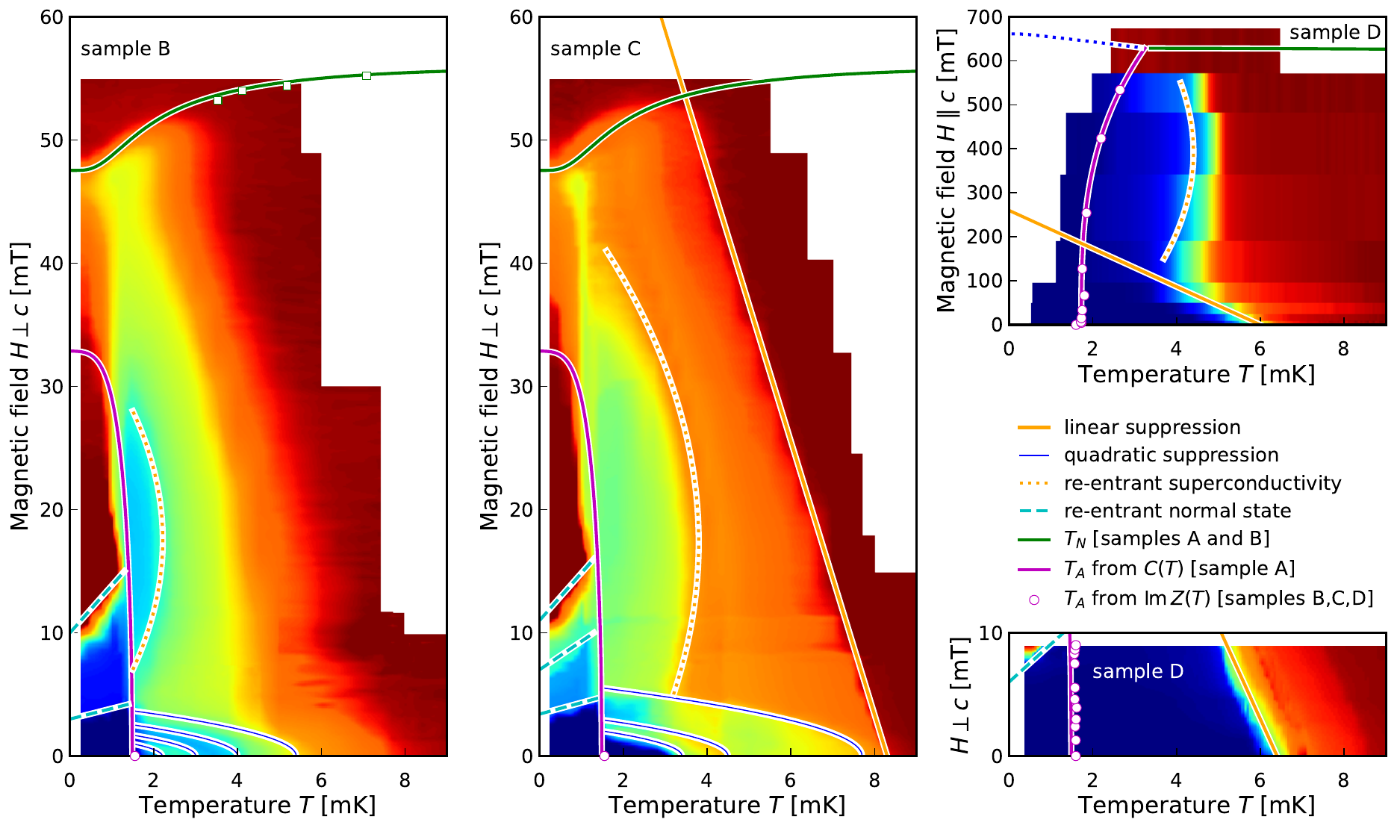}}
\caption{Contours of superconducting phases on the $\Re Z(T,H)$ color maps
corresponding to schematic diagrams Figs.~1G (linear, i.e. orbital, suppression),
1H (quadratic Pauli suppression), 1I (re-entrant superconductivity) and 1J (re-entrant normal state).
Here the linear suppression is extrapolated to $T = 0$ as a straight line.
Typical curvature of lines of orbital suppression is illustrated
in Fig.~\ref{fig:Maki}}\label{fig:SC:contours}

\vskip1em

\centerline{\includegraphics[scale=0.47]{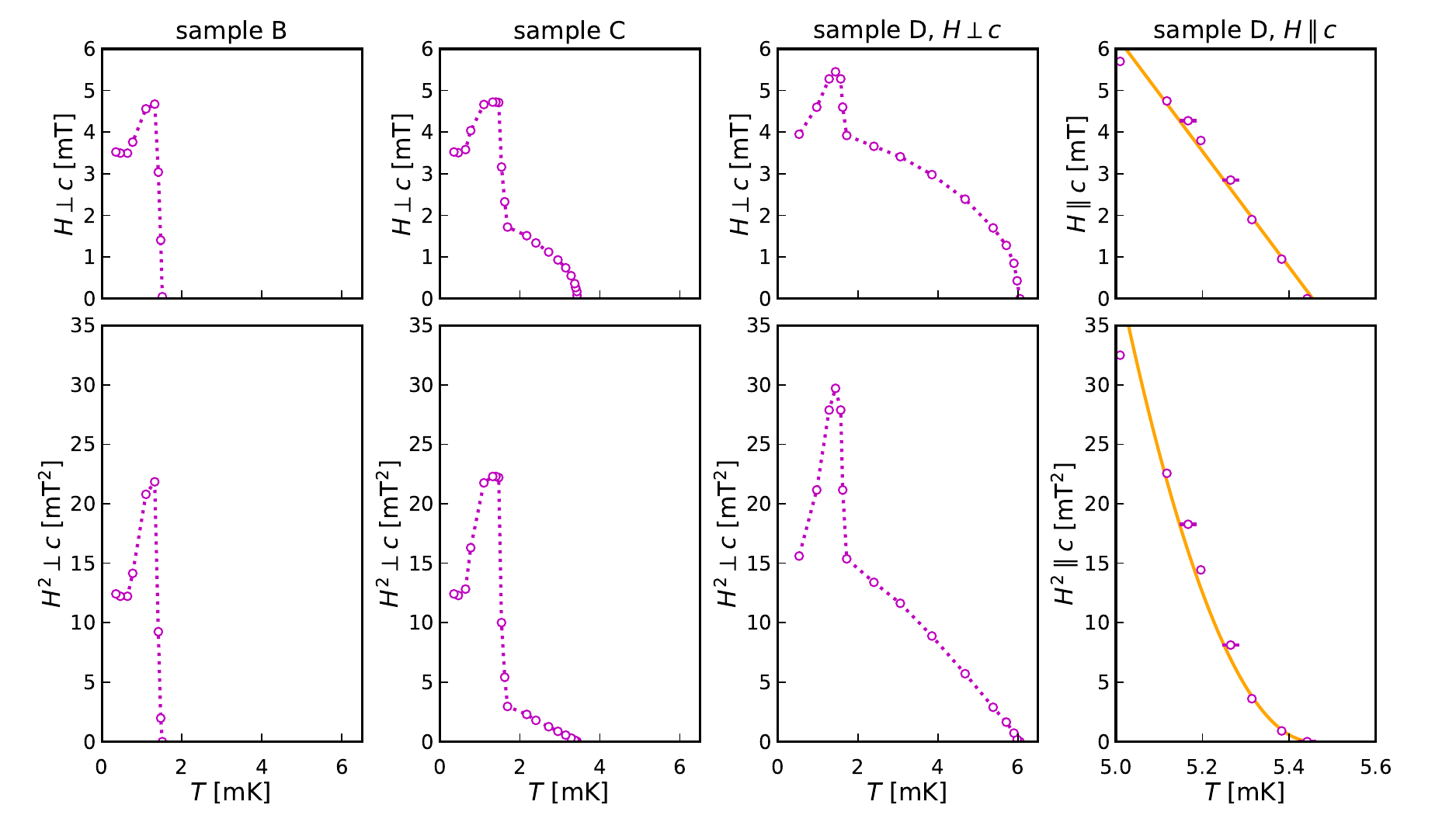}}
\caption{Onset of flux quantization in the three samples. For in-plane fields the contours
are parabolic above $T_A = 1.5$\,mK, except sample B, where flux quantization is only observed
below $T_A$. For $H \parallel c$ at low fields the suppression is linear;
here the flux quantization is limited by the superconducting transition in aluminium wires connected to the sample at 6\,mK. Their critical field is suppressed from $\mu_0 H_c^{\text{Al}} = 10.5$\,mT
due to the demagnetizing factor $\sim 0.5$ of a cylindrical wire in perpendicular field.}\label{fig:fluxqcontours}
\end{figure*}

The heat capacity signatures of the critical field $H_N$ of AFM1 were limited to $T > 30$\,mK.
A signature in magnetoresistance, see Fig.~\ref{fig:MR}, allows us to extend
the measurements of $H_N(T)$ down to 4\,mK, below which the superconductivity
starts to dominate the field dependence of the sample impedance.
Additionally, a magnetocaloric signature (kink in $T_{S = \mathrm{const}}(B)$ line) has been observed at 1\,mK.

While $H_N(T)$ monotonically grows on cooling down to 20\,mK, at lower temperatures
it exhibits positive slope, see Fig.~\ref{fig:AFM1}, unusual for an antiferromagnet.
This effect can be quantitatively accounted for by the hyperfine coupling
at $^{171}$Yb and $^{173}$Yb sites. This interaction
$$
\mathcal{H}_{\text{hf}} = g\mu_N A_{\text{hf}} \,\vec{\widehat{I}}\cdot \vec\mu_e
= -\vec B_{\text{hf}} \cdot (g\mu_N \vec{\widehat{I}}) = - \vec b_{\text{hf}} \cdot \vec\mu_e
$$
can be viewed as an effective magnetic field
$$
\vec B_{\text{hf}} = -A_{\text{hf}} \,\vec\mu_e
$$
exerted by the electronic moment $\vec\mu_e$ on the nuclear magnetic moment
$g\mu_N \widehat{\vec I}$,
or a reciprocal effective field
$$
\vec b_{\text{hf}} = -g\mu_N A_{\text{hf}} \,\vec{\widehat{I}}
$$
exerted by the nuclear moment on the magnetic moment of the electron~\cite{Knapp2023}.
Here $A_{\text{hf}} = 102\,\mathrm{T}/\mu_B$ is the hyperfine constant, $g$
is the nuclear $g$-factor, $\vec{\widehat{I}}$ is the nuclear spin operator.

The application of $\sim 60\,\mathrm{mT}\perp c$ to YRS{} polarizes the Yb electrons
to approximately $\mu_e = 0.1\mu_B$, where $\mu_B$ is Bohr's magneton~\cite{Knapp2023}.
This electronic polarization splits the $^{171}$Yb and $^{173}$Yb
nuclear energy levels, leading to a temperature-dependent nuclear polarization.
In case of $^{173}$Yb ($I=5/2$) the nuclear Hamiltonian
\begin{equation}\label{eq:H173}
\mathcal{H}_{173} = g\mu_N A_{\text{hf}} \, \vec{\widehat{I}}\cdot \vec\mu_e +
\frac{e^2qQ}{4I(2I-1)}(3\widehat{I}_c^2 - I(I+1))
\end{equation}
includes quadrupolar interaction and requires numerical diagonalization for $\vec\mu_e \perp c$.
The simpler case of $^{171}$Yb ($I=1/2$) can be treated analytically (Schottky model).
After the diagonalization we calculate the thermal average of $\vec{\widehat{I}}$
and the effective field $\vec b_{\text{hf}}$, which lies parallel to $\vec\mu_e$
and thus to the external field $\vec H$. The temperature dependence
for the two magnetically active Yb isotopes and spatial average $\langle b_{\text{hf}}\rangle$
for the natural abundance of Yb (14\% $^{171}$Yb, 16\% $^{173}$Yb, 70\% non-magnetic isotopes)
are shown in Fig.~\ref{fig:AFM1}B.
Adding this spatially averaged $\langle b_{\text{hf}}\rangle$ to
the measured critical field $\mu_0 H_N$ of AFM1 leads to a conventional $H_N^*(T)$ with no temperature dependence below 10\,mK.

Ignoring the difference between $\vec B$ and $\mu_0 \vec H$ in the weakly-magnetic \YRS{}
we can write the Zeeman and hyperfine interactions for Yb electrons
$$
\mathcal{H} = -\mu_0 \vec H \cdot \vec\mu_e + \mathcal{H}_{\text{hf}}
= - (\mu_0 \vec H + \vec b_{\text{hf}}) \cdot \vec \mu_e = -\mu_0 \vec H^* \cdot \vec \mu_e
$$
as Zeeman interaction alone with effective field
\begin{equation}
\mu_0 \vec H^* = \mu_0 \vec H + \vec b_{\text{hf}}.
\end{equation}
Thus $H_N^*$ is the spatially averaged effective field $H^*$ at the AFM1/PM
phase boundary. The non-monotonic $H_N(T)$ is thus quantitatively taken into account by making two assumptions:
\begin{itemize}
\item The AFM1 state is driven by 4f electrons of Yb, which experience effective hyperfine field
in addition to the external magnetic field.
\item The polarised electronic moments of the strongly-correlated AFM1 state
are uniform despite spatially varying hyperfine field.
\end{itemize}
To self-consistently include the temperature dependence of $H_N$ we approximate the
experimentally observed field dependence of $\mu_e$~\cite{Knapp2023} with a temperature-independent
$$
\mu_e = 0.15\mu_B / \mathrm{T} \times \big(\mu_0 H + \langle b_{\text{hf}}\rangle\big)
$$
when calculating $\langle b_{\text{hf}}\rangle$.
The experimental $H^*_N(T)$ data are fitted by a polynomial
\begin{align}
\big(\mu_0 H_N^*(T)\big)^2 = & \big(\mu_0 H_N(T) + \langle b_{\text{hf}}\rangle\big)^2 \notag\\
= & -0.005006\,\mathrm{mT^2/mK^3} \times T^3\notag\\
& - 0.3110\,\mathrm{mT^2/mK^2} \times T^2 \notag\\
& + 0.4281\,\mathrm{mT/mK} \times T \notag\\
& + 3271\,\mathrm{mT}^2.
\end{align}
This equation determines the AFM1/PM phase boundary shown with a solid green line in Figs.~1, \ref{fig:AFM1}A and \ref{fig:SC:contours}.
The precise choice of $\mu_e$ at $H_N$ only weakly affects the model.
Importantly $\langle b_{\text{hf}}(T=0) \rangle$ is a weak function of $\mu_e$.
See Ref.~\cite{Knapp2025} for a more complete analysis of this phase boundary taking into account the field-dependent magnetic susceptibility 

\subsubsection{Out-of-plane magnetic phase diagram}

To estimate the out-of-plane critical field of AFM1 we recall
the anisotropy $H_{N \parallel c} = 11 H_{N \perp c}$~\cite{Gegenwart2002} observed above 20\,mK.
In line with the assumptions made in the previous paragraph, at lower temperatures the magnetic
fields in this relationship are replaced with the effective fields. Then,
assuming no temperature dependence to the anisotropy factor,
$\mu_0 H_{N \parallel c}^{\text{est}} = 11\mu_0 H_N^* - \langle b_{\text{hf}}\rangle$,
where $H_N^*$ denotes the in-plane effective critical field, as before.

Even when fully polarized $^{171}$Yb and $^{173}$Yb the averaged hyperfine field
$\langle b_{\text{hf}}\rangle\sim 10$\,mT is insignificant on the scale of out-of-plane critical field $H_{N \parallel c}\sim 600$\,mT. Furthermore, we expect $b_{\text{hf}}$ to develop below 1\,mK,
see Fig.~\ref{fig:AFM1}D, due to small electron moments $\mu_e \sim 0.01\mu_B$ at $H_N$ estimated from the magnetic susceptibility measurements~\cite{Gegenwart2002}.
Thus, in Fig.~3 we estimate the critical field of AFM1 as
\begin{equation}
H_{N}^{\text{(est)}} = 11H_N^*.
\end{equation}
Overall the magnetic phase diagram of \YRS, shown in Figs.~1A and 3, resembles
that of Yb$($Rh$_{0.82}$Co$_{0.18})_2$Si$_2$~\cite{Hamann2019}.
The relationship between the magnetic structure of the electro-nuclear order of the AFM2 phase in \YRS{}
and the eponymous phase in Ref.~\cite{Hamann2019}, remains an open question.

\subsection{Contours on superconducting phase diagrams}

Figs.~1G-J schematically illustrate various contours of constant resistance observed in
samples B, C and D, attributed to distinct superconducting phases. One type of contour,
the parabolic Pauli-limited contours (Fig.~1H) are overlayed on the resistance maps in Fig.~2.
The same is shown for all contour types in Fig.~\ref{fig:SC:contours}.
Complementary contours marking the onset of flux quantization are shown in Fig.~\ref{fig:fluxqcontours}.

\begin{figure}[b!]
\centerline{\includegraphics[scale=0.47]{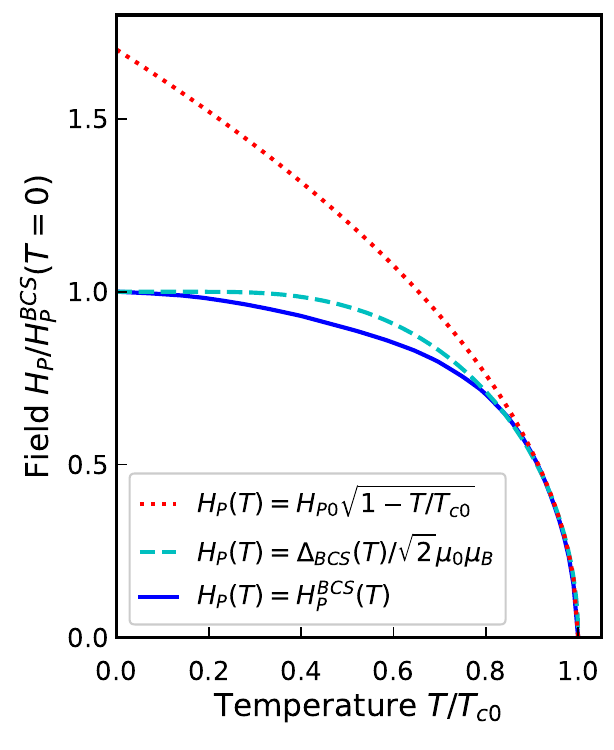}}
\caption{\label{fig:HPauli}Temperature dependence of Pauli limiting field.
The numerical calculation $H_P^{BCS}(T)$ within the BCS theory for an s-wave superconductor~\cite{Sarma1963}
closely follows the temperature dependence of the s-wave BCS energy gap $\Delta^{BCS}(T)$.
The Ginzburg-Landau expression $H_P(T) = H_P(0) \sqrt{1 - T/T_{c0}}$ agrees with the numerical results
close to $T_c$.}
\end{figure}

\subsection{Temperature dependence of~the~Pauli-limited~field}

Figure~\ref{fig:HPauli} illustrates the temperature dependence of the Pauli-limited field
for an s-wave BCS superconductor with a $g$ factor of 2.
A numerical calculation, considering the spin susceptibility in the presence
of pairing and the gradual suppression of the energy gap by field with both 1st and 2nd order phase transitions~\cite{Sarma1963} is closely approximated by a simple expression
$
H_P(T) = \Delta(T) / \mu_0 \mu_B \sqrt{2}
$
obtained from the exact $T = 0$ result~\cite{Sarma1963}
by replacing $H_P(0)$ and $\Delta(0)$ with their temperature-dependent values.
Note that here $\Delta(0)$ and $\Delta(T)$ represent the gap at zero field,
in the absence of Pauli suppression.

Both the temperature dependence of the energy gap and the Pauli limiting field are affected by the presence of nodes in the energy gap. The observed nearly-perfect parabolas, see Figs.~2 and \ref{fig:fluxqcontours}, show stronger temperature dependence below $T = 0.5T_{c0}$ than the BCS model of an isotropic fully-gapped s-wave superconductor, Fig.~\ref{fig:HPauli}. This may point to strong gap variations over the Fermi surface.


\subsection{Anisotropy of the Pauli-limiting field}

In the AFM1 phase with in-plane magnetic fields, Fig.~2,
we observe parabolas with $H_P = 2$-6\,mT in the $T \to 0$ limit. 
Scaling this by the $g_{ab} / g_c = 20$ factor, we obtain 40-120\,mT$\,\,\parallel c$.

In the AFM2 phase the in-plane critical fields, Figs.~1 and 2, drop from 5-15\,mT just below $T_A$
to 3-10\,mT at the lowest temperature. Applying the $g_{ab} / g_c$ scaling, we obtain 60-300\,mT$\,\,\parallel c$.

\subsection{Pauli vs orbital limit, coherence length, Maki~parameter}

\begin{figure}[t!]
\centerline{\includegraphics[scale=0.47]{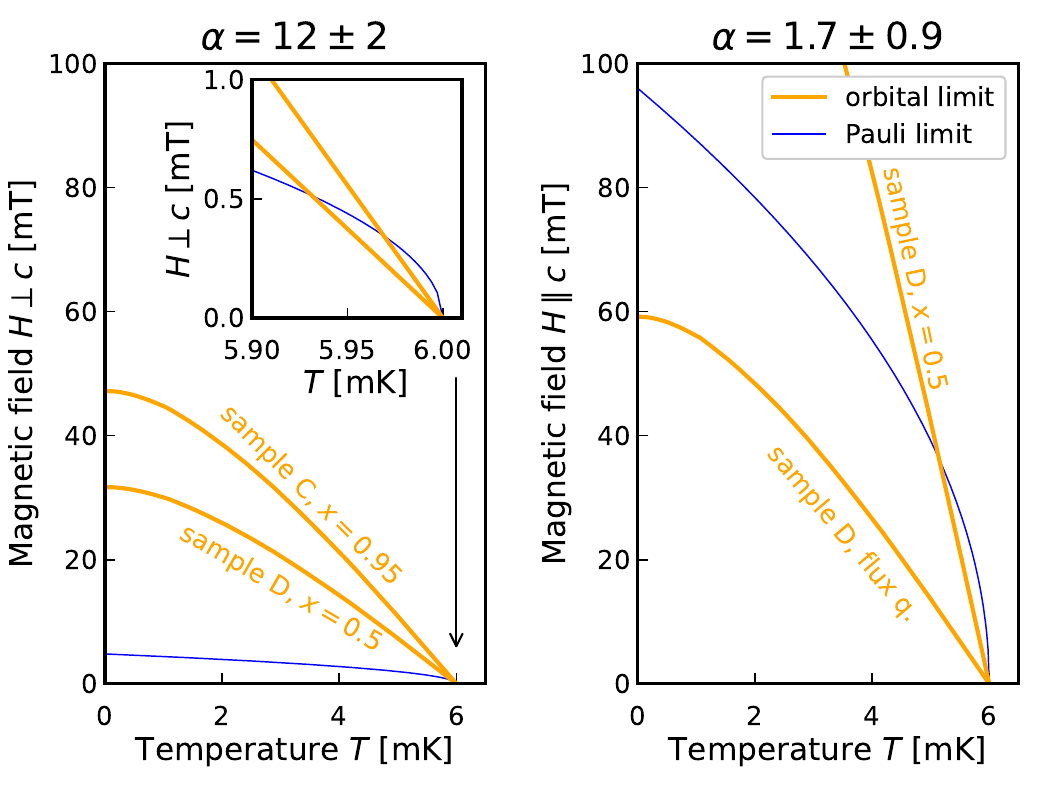}}
\caption{Pauli and orbital limiting fields for two field orientations
and corresponding Maki parameter $\alpha$.
The experimental signatures are rescaled to a typical $T_{c0} = 6$\,mK to allow direct comparison
between them. The rescaling assumes $H_{c2}^{\text{orbital}} \propto T_{c0}$, preserving the $\partial T_c/\partial H$ slope.
The lines of linear suppression are extrapolated following
the purely orbital model of Werthamer, Helfand and Hohenberg,
the $\alpha = 0, \lambda_{SO} = 0$ curve in Fig.~3 in Ref~\cite{Werthamer1966}.
The lines derived from $Z(T,H)$ are denoted by the $x = \Re Z / R_0$ ratio.
With in-plane fields the orbital limit only dominates the Pauli limit
within less than tens of $\mu$K below $T_{c0}$, resulting in clearly observed parabolas characteristic
of the Pauli limit. This temperature window is significantly
wider and may extend all the way from $T_{c0}$ to $T = 0$
for the out-of-plane field.
The flux quantization curve is derived from Fig.~\ref{fig:fluxqcontours}.}\label{fig:Maki}
\end{figure}

We attribute the conventional linear suppression of superconductivity with field, Fig.~1G,
to orbital effects and estimate the (upper) critical field using Werthamer-Helfand-Hohenberg
calculation~\cite{Werthamer1966}	
\begin{equation}
H_{c2}^{\text{orbital}}(T = 0) \approx 0.7 T_{c0} \bigg / \left(\frac{\partial T_c}{\partial H}\right)_{H=0}.
\end{equation}
Assuming isotropic coherence length $\xi_{ab}$ in the $ab$ plane, and a different value $\xi_c$
along the $c$ axis, we obtain the $T=0$ coherence lengths from
\begin{equation}\label{eq:xi}
\begin{array}{rcl}
\xi_{ab}^2 &=& \Phi_0 \big/ {2\pi \mu_0 H_{c2 \parallel c}^{\text{orbital}}},\\
\xi_{ab} \xi_c &=& \Phi_0 \big/ {2\pi \mu_0 H_{c2 \perp c}^{\text{orbital}}}.
\end{array}
\end{equation}
The initial slope ${\partial T_c(H)} / {\partial H}$ of the signatures
of superconductivity that are linearly suppressed by field, Fig.~\ref{fig:SC:contours},
is less reproducible than the $H_P / T_{c0}$ ratio characterizing the Pauli suppression, Fig.~2, showing variations
not only from one sample to another, but also between different signatures
within the same sample.

From the initial slopes of the $\Re Z / R_0 = 0.5$ contours in sample D alone, measured for two field orientations,
$H_{c2 \parallel c}^{\text{orbital}}(T = 0) = 183$\,mT,
$H_{c2 \perp c}^{\text{orbital}}(T = 0) = 34$\,mT, leading to $\xi_{ab} = 42$\,nm, $\xi_c = 235$\,nm.
Including the initial slopes of other observed signatures in the analysis, we estimate
$H_{c2 \parallel c}^{\text{orbital}}(T = 0) = 121 \pm 62$\,mT,
$H_{c2 \perp c}^{\text{orbital}}(T = 0) = 39\pm16$\,mT,
$\xi_{ab} = 56 \pm 13$\,nm, $\xi_c = 125 \pm 82$\,nm,
$\xi_c / \xi_{ab} = 3 \pm 2$. The order of magnitude of both $\xi_{ab}$ and $\xi_c$ is 100\,nm,
as quoted in the main text.

The above analysis of the initial slopes effectively extrapolates the behavior in small fields.
However Fig.~3 shows superconductivity exceeding the $H_{c2 \parallel c}^{\text{orbital}}(T = 0) = 183$\,mT, quoted above. At $H = H_N$ the superconductivity is destroyed by the change in the underlying magnetism, therefore $H_{c2} > H_N$. This leads to an upper bound
$\xi_{ab} < \sqrt{\Phi_0 / {2\mu_0 H_{N\parallel c}}} = 23$\,nm for the high-field superconductivity.
We can put no such bound on $\xi_c$, as it is convoluted with $\xi_{ab}$
in Eq.~\eqref{eq:xi}.

The relative strength of the orbital and Pauli suppression of superconductivity with magnetic field
can be parameterized via the Maki parameter~\cite{Maki1964,Maki1966}
$$
\alpha = \frac{H_{c2}^{\text{orbital}}(T=0)\sqrt{2}}{H_P(T=0)}.
$$
Figure~\ref{fig:Maki} shows both linear and quadratic signatures rescaled to a typical $T_{c0} = 6$\,mK.
For $H \perp c$ we obtain large $\alpha = 12 \pm 2$, that enables the Pauli suppression to dominate the signatures except within tens of $\mu$K of $T_{c0}$.
For $H \parallel c$ we find moderate $\alpha = 1.7 \pm 0.9$,
for which the orbital suppression prevents the observation of the parabolas.

\clearpage

\section{Candidate superconducting order~parameters}

For the $D_{4h}$ symmetry the list of candidate order parameters is summarized in Table~\ref{tab:D4h}.
In general an order parameter corresponding to a particular IR is a linear combination
of an easy plane ($\vec d \perp c$) and an easy axis ($\vec d \parallel c$) block, i.e.
\begin{equation}
\vec d(\vec k) = \Delta_{ab} (k_a\hvec a  + k_b\hvec b) + \Delta_c k_c \hvec c
\end{equation}
for the $A_{1u}$ IR.
Below we only consider order parameters with either $\Delta_c = 0$ or $\Delta_{ab} = 0$,
selected by the spin-orbit coupling.

Four of the candidate states are examples of the \emph{helical} phase.
A general helical order parameter can be written as
\begin{equation}\label{eq:d:helical}
\vec d(\vec k) = \Delta \mtrx R \big(\vec k - \hvec c (\vec k \cdot \hvec c)\big),
\end{equation}
where $\mtrx R$ is a rotation matrix.
These states have identical gap structure with point nodes at $\vec k \parallel c$
or no nodes, depending on the Fermi surface topology.

Three other states listed in Table~\ref{tab:D4h} are \emph{nematic}, where
\begin{equation}
\vec d(\vec k) = \Delta \hvec d (\vec k \cdot \hvec p).
\end{equation}
Here $\hvec p$ denotes the direction of gap maximum, with a line node at $\vec k \perp\hvec p$.

\subsection{Pauli limit in easy plane ($\vec d \perp c$) states}

In this section we derive the Pauli limiting field of various easy-plane order parameters
from Table~\ref{tab:D4h}.
We consider the paramagnetic coupling
\begin{equation}\label{eq:dH}
F_H = - \frac{1}{2}\chi_{\mu\nu}\{\vec d(\vec k)\} H_{\mu} H_{\nu},
\end{equation}
where $\chi_{\mu\nu}\{\vec d(\vec k)\}$ is the tensor of magnetic susceptibility
in a superconducting state with order parameter $\vec d(\vec k)$.
Subtracting the normal-state paramagnetic energy $-\chi^{\text N}_{\mu\nu}H_{\mu}H_{\nu} / 2$
and expanding in powers of $\vec d$ and $\vec d^*$,
Eq.~\eqref{eq:dH} can be rewritten as
\begin{equation}\label{eq:dH:GL}
F_H = \tilde\alpha \big\langle (\vec d^* \cdot \vec H)(\vec d \cdot \vec H)\big\rangle\FS,
\end{equation}
similar to Eq.~(5.46) in Ref.~\cite{VWbook}.
The coefficient $\tilde\alpha$ depends on $T_{c0}$ and various normal-state properties,
including the $g$ factor, therefore we must consider separate values of $\tilde\alpha$ 
for $\vec H \parallel c$ and $\vec H \perp c$.
This model does not take into account the potential variation of the $g$ factor
over the Fermi surface invoked in  the pseudospin picture~\cite{Sauls1994}.


The different easy-plane states are distinguished from each other
by the in-plane anisotropy of the Pauli limiting field.
Their common property, is the independence of $F_H$ on
the out-of-plane field $H_c$, and therefore the absence of Pauli limit
for $\vec H \parallel c$. This contrasts the easy-axis states
examined in Sec.~\ref{sec:easyaxis} below.

\begin{table}[t!]
\caption{Candidate triplet order parameter states
corresponding to the irreducible representations (IRs) of $D_{4h}$ group%
~\cite{Mineev1999,Yip1993,Aoki2022}.
For the 2D IR $E_u$ we first show the general form of the order parameter
and then list the stable states found using Ginzburg-Landau analysis~\cite{Mineev1999}.
Spin-orbit coupling is expected to select either $\vec d \perp c$ or $\vec d \parallel c$ block.
In the $\vec d \perp c$ block of $A_{2u}$ we only retain the term linear in $k$
and drop the additional term of the 3rd order in $k$ discussed in Refs.~\cite{Mineev1999,Volovik1985}.
The superconducting phases discussed in detail are labelled in brackets.%
}\label{tab:D4h}
\medskip

\centerline{\begin{tabular}{@{~}l@{~~~~}ll@{~~~~}l@{~}}
\hline
\rule{0pt}{1.1em}IR & $\vec d \perp c$
& $\vec d \parallel c$ \\
\hline
\rule{0pt}{1.1em}%
$A_{1u}$ & $k_a\hvec a + k_b\hvec b$ ~~(helical) & $k_c\hvec c$ ~~~~~~~~~~~~~(nematic) \\
$A_{2u}$ & $k_a\hvec b - k_b\hvec a$ ~~(helical) & $k_a k_b (k_a^2-k_b^2) k_c \hvec c $ \\
$B_{1u}$ & $k_a\hvec a - k_b\hvec b$ ~~(helical) & $(k_a^2 - k_b^2) k_c \hvec c$ \\
$B_{2u}$ & $k_a\hvec b + k_b\hvec a$ ~~(helical) & $k_a k_b k_c \hvec c$ \\
$E_{u}$  & $\eta_1 k_c\hvec a + \eta_2 k_c\hvec b$ & $\eta_1 k_a\hvec c + \eta_2 k_b\hvec c$\\[0.1em]
\hline
\rule{0pt}{1.1em}%
$E_u$ & $k_c\hvec a, k_c\hvec b$ ~~~~~(nematic) & $k_a\hvec c, k_b\hvec c$ \,~~~~~~(nematic)\\
$E_u$ & $k_c (\hvec a \pm \hvec b)$ \,~~(nematic) & $(k_a \pm k_b) \hvec c$ \,~~(nematic)\\
$E_u$ &  $k_c (\hvec a \pm i\hvec b)$ ~(spin chiral)~~ & $(k_a \pm ik_b)\hvec c$ ~~(chiral)\\
\hline
\end{tabular}}
\end{table}

\subsubsection{$A_{1u}, A_{2u}, B_{1u}, B_{2u}$: helical states}


For the helical states corresponding to the four 1d IRs the magnetic energy
$F_H = \tilde\alpha |\Delta|^2 \langle k_a^2 \rangle\FS H^2_{ab}$
is determined by the magnitude of the in-plane field $H_{ab} = \sqrt{H_a^2 + H_b^2}$.
Here the gap amplitude $\Delta$ has the same meaning as in Eq.~\eqref{eq:d:helical}
and the derivation relies on tetragonal symmetry: $\langle k_a^2 \rangle\FS = \langle k_b^2 \rangle\FS$.
By including $F_H$ in a simple Ginzburg-Landau model
$$
F = \alpha (T  - T_{c0})|\Delta|^2 + \frac{u}{2}|\Delta|^4 + F_H,
$$
we arrive at quadratic Pauli suppression of the transition temperature with
in-plane field, independent of its orientation within the plane,
$$
T_c(H) = T_{c0} - \frac{\tilde\alpha}{\alpha} \langle k_a^2 \rangle\FS H_{ab}^2.
$$
We note that this isotropic in-plane Pauli limit is not linked to the isotropic in-plane
energy gap, $|\vec d(\vec k)| = |\Delta|$ for all $\vec k \perp c$, exhibited by
the model helical states listed in Table~\ref{tab:D4h}.
The latter property is not required by the symmetry arguments and
the order parameters can have a more general form~\cite{Yip1993}.
For example, for the IR $A_{2u}$ we can consider 
$\vec d(\vec k) = \Delta (k_a^3\hvec b - k_b^3\hvec a)$, that
exhibits in-plane gap variations between $|\Delta|$ at $\vec k \parallel [100]$ and $[010]$
and $|\Delta/2|$ at $\vec k \parallel [110]$ and $[1\bar10]$ momentum directions.
Nevertheless the magnetic energy
$F_H = \tilde\alpha |\Delta|^2 \big\langle k_a^6 \big\rangle\FS H_{ab}^2$
only depends on the absolute value of the in-plane field, as for the simpler
$\vec d(\vec k) = \Delta (k_a\hvec b - k_b\hvec a)$.


We propose the helical phase to be the superconducting state that exhibits
the observed in-plane Pauli limit with reproducible $H_P / T_{c0}$ ratio, Fig.~2.
Each version of the helical phase corresponds to a 1D IR.
Therefore crossing the Pauli-limiting field will likely lead to a
transition to the normal state, accounting for the abrupt jump in $\Re Z$ manifested
at some of the Pauli-limited signatures in Fig.~2.
We note that a field-induced transition to another IR is also possible.
 
\subsubsection{$E_u$: easy-plane nematic and spin chiral states}%

The remaining IR $E_u$ is two-dimensional,
allowing for a manifold of superconducting states, all having the same $T_{c0}$
but different pairing energy.
The general form of the order parameters can be written as %
$
\vec d(\vec k) = \eta_1 k_c\hvec a + \eta_2 k_c\hvec b.
$
We can rewrite this as
\begin{equation}\label{eq:dEu}
\vec d(\vec k) = \Delta_+ e^{i(\phi - \theta)} k_c \frac{\hvec a + i\hvec b}{2}
+ \Delta_- e^{i(\phi + \theta)} k_c \frac{\hvec a - i\hvec b}{2}
\end{equation}
in terms of the pairing amplitudes $\Delta_+$ and $\Delta_-$ for the $\uparrow\uparrow$ and
$\downarrow\downarrow$ Cooper pairs~(see Eq.~(3.34) in Ref.~\cite{VWbook}).
Here $\Delta_+$ and $\Delta_-$ are real and non-negative,
and the phases of the order parameter components are written down explicitly.

In a magnetic field $\vec H = H_{ab} (\hvec a \cos\mu + \hvec b \sin\mu)$
of arbitrary in-plane direction defined by the azimuth $\mu$,
the magnetic energy Eq.~\eqref{eq:dH:GL} takes form
\begin{equation}\label{eq:dH:Eu}
F_H = \frac{1}{4}\tilde\alpha \langle k_c^2\rangle\FS H_{ab}^2 \big[\Delta_+^2 + \Delta_-^2
+ 2\Delta_+ \Delta_- \cos 2(\theta - \mu)\big].
\end{equation}
Importantly, for any $\mu$ there exists a state with no Pauli limit, $F_H = 0$.
From Eq.~\eqref{eq:dH:Eu} we conclude that this state has equal amount of
$\uparrow\uparrow$ and $\downarrow\downarrow$ pairs ($\Delta_+ = \Delta_-$)
and $\theta = \mu \pm \pi/2$. This is a \textbf{nematic state}
\begin{equation}\label{eq:d:nematic}
\vec d(\vec k) = \Delta k_c (\hvec a \cos\theta + \hvec b \sin\theta) = \Delta k_c \hvec d
\end{equation}
with spin director $\hvec d = \hvec a \cos\theta + \hvec b \sin\theta$
and amplitude $\Delta = \Delta_+ e^{i\phi}$.
To avoid Pauli limit, it must align $\hvec d \perp \vec H$.

In zero magnetic field the Ginzburg-Landau theory of two-component order parameters
predicts one of three doubly-degenerate ground states, listed at the bottom of Table~\ref{tab:D4h}:
nematic state with $\theta = 0$ or $\theta = \pi/2$;
nematic state with $\theta = \pm \pi /4$; or \textbf{spin chiral state},
where only one of $\Delta_+$ or $\Delta_-$ is non-zero~\cite{Mineev1999,Volovik1985}.
Our numerical analysis of this model suggests that each of these grounds states
continuously evolves into the Pauli-unlimited nematic state upon application
of in-plane field comparable to the Pauli limiting field of the helical states.
This transformation may reduce the pairing energy and gap amplitude, but importantly
$T_{c0}$ remains unchanged.
If the zero field ground state is nematic, all the field does
is rotate $\hvec d$ towards an in-plane direction perpendicular to the field.
For the spin chiral state the intermediate order parameters on the way to nematic state have
non-zero and unequal $\Delta_+$ and $\Delta_-$, similar to the $\beta$-P$_2$-polar evolution in \He3 in nematic aerogel~\cite{Surovtsev2019}.

The spin chiral state leaves either the (pseudo)spin down or (pseudo)spin up Fermi surface completely unpaired, therefore it is unlikely to be the ground state\footnote{In \He3 the analogue of the spin chiral state is the $\beta$ or P$_1$ phase, that requires strong magnetic field to be stable~\cite{Surovtsev2019,Dmitriev2021}.
By the same mechanism in YbRh$_2$Si$_2$ the nematic state may evolve
into the spin chiral state in strong magnetic field $\vec H \parallel c$, but it is unlikely to be relevant to the low-field part of the phase diagram discussed here.
Spin-polarized states have also been considered in ferromagnetic superconductors~\cite{Mineev1999}.}.
Nevertheless it is important to consider the spin chiral state, because this order parameter
has isotropic in-plane Pauli limit, the defining property of the helical states. 
Nevertheless, our identification of the helical states remains unaffected by this coincidence,
because in the case of spin chiral state the Pauli-limited suppression of the superconductivity is precluded by the continuous transformation into the nematic state.

Thus we conclude that the regions of the sample that support the easy-plane $E_u$ IR would exhibit the nematic superconducting state Eq.~\eqref{eq:d:nematic} beyond the in-plane Pauli limit. Experimentally establishing the evolution of the order parameter at low fields and the influence of symmetry breaking by strain and antiferromagnetism is beyond the scope of this work.

\subsection{Pauli limit in easy axis ($\vec d \parallel c$) states}\label{sec:easyaxis}

\begin{figure}[b!]
\centerline{\includegraphics[scale=0.47]{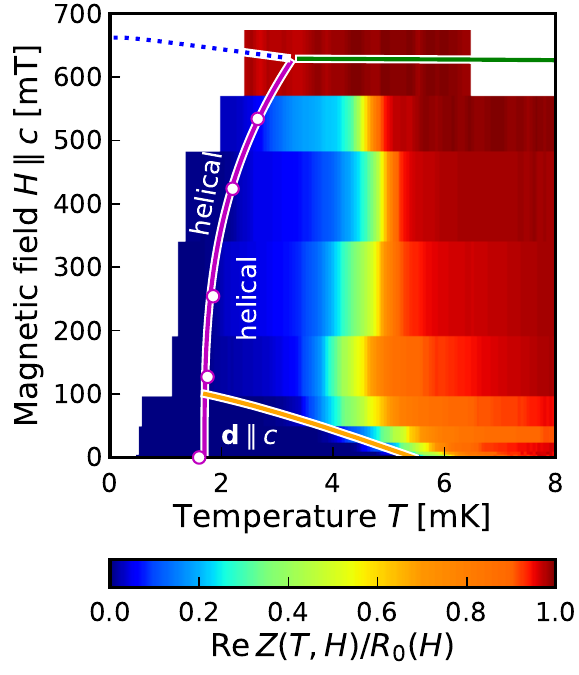}}
\caption{\label{fig:helical:chiral}
A kink in $T_c(H)$ observed for out-of-plane fields can be interpreted as a phase transition
between two superconducting phases, each present in sufficient fraction
of the sample to result in $\Re Z = 0$.
In this picture the critical field of the low-field phase is close to the Pauli limit estimated
from the $g$-factor anisotropy (see main text), therefore the low-field phase
may have an easy-axis ($\vec d \parallel c$) order parameter, Pauli-limited for this field orientation.
In this model the easy-axis state complements the helical state, instead of the
easy-plane nematic state.}
\end{figure}

For all easy-axis states listed in Table~\ref{tab:D4h} the magnetic energy Eq.~\eqref{eq:dH}
$F_H = \tilde\alpha \langle \vec d^* \cdot \vec d \rangle\FS H_c^2$
depends on the out-of-plane field $H_c$, but not on $H_{ab}$.
Therefore these states have similar magnetic properties: Pauli limit for $\vec H \parallel c$ and no Pauli limit for any in-plane field orientation.
Due to the lack of in-plane Pauli limit any of these order parameters should be considered as an alternative to the easy-plane nematic state.

We observe no clear parabolic contours for $\vec H \parallel c$,
that would unambiguously demonstrate the Pauli limit of an easy-axis state.
However, the initial slopes of the low-field signatures, Figs.~3, \ref{fig:fluxqcontours}, \ref{fig:Maki}, characterized by a relatively small Maki parameter $\alpha \sim 1$, are compatible
with $\vec d \parallel c$ states, as illustrated in Fig.~\ref{fig:helical:chiral}.

A strong candidate of this type is the chiral state $\vec d(\vec k) = \Delta (k_a \pm ik_b)\hvec c$.
This state has the same gap structure $|\vec d(\vec k)|$ as the helical phase,
with point nodes or fully gapped. In the weak coupling limit these states are favored
over all other candidates from Table~\ref{tab:D4h}, which have line nodes and thus leave more electrons unpaired.

\subsection{Accidental degeneracy of irreducible representations}

Each IR is in general characterized by its own critical temperature, and
so far we considered different IRs one by one.
Accidental degeneracy between different IRs has been invoked
in UTe\textsubscript{2} to create a chiral phase by combining
order parameters corresponding to different IRs~\cite{Aoki2022}.

In the heterogeneous \YRS{} crystal the overall critical temperature $T_{c0}$,
equal to the highest critical temperature of any IR, varies
from one region of the crystal to another, driven by the nonuniformity in the crystal.
It is natural that the critical temperature for any IR does the same, unless this IR is completely
unstable.
There is no reason why different IRs would respond to the nonuniformities in identical way,
therefore, while the degeneracy is very likely to occur, we expect it to be restricted
to minority regions of the crystal, i.e. at the boundaries between domains
where two different IRs have the highest critical temperature.
Thus the robust observations of in-plane Pauli limit and superconductivity beyond it must
find explanation without the need to mix order parameters corresponding to different IRs.
This justifies our approach to examine different IRs one by one.

\subsection[Josephson coupling to a conventional superconductor]{Josephson coupling
between triplet states in\\\YRS{} and a conventional superconductor}

R.~Jin \emph{et al.}~\cite{Jin2000} propose that the critical current of
a Josephson junction formed by a triplet and singlet superconductors is proportional to
\begin{equation}\label{eq:I0Jin}
I_c \propto \langle \vec d(\vec k) \cdot \vec k \times \hvec n\rangle\FS,
\end{equation}
where $\vec n$ is the direction normal to the interface.
Analyzing the candidate states from Table~\ref{tab:D4h}
in terms of Eq.~\eqref{eq:I0Jin} for $\vec n \parallel c$
we find that the $A_{2u}$ helical phase the only state for which $I_c \ne 0$
in case of unperturbed $D_{4h}$ symmetry ($\langle k_a^2 \rangle = \langle k_b^2 \rangle$,
$\langle k_a k_b \rangle = \langle k_a k_c \rangle = \langle k_b k_c \rangle = 0$).

The flux quantization, Figs.~2, \ref{fig:fluxq}, requires a continuous
phase-coherent superconducting current in a loop comprised by the \YRS{} sample,
which we argue is a triplet superconductor, and conventional s-wave superconductors Al, Nb and NbTi.
In this loop the connections to the sample are made by Al wires ultrasonically bonded to the
surface perpendicular to the $c$ axis. An ideal point contact between \YRS{} and Al
would form a Josephson junction; in this case the observation of flux quantization is a signature
of the $A_{2u}$ helical phase.
The microscopic structure of the \YRS{}-Al wirebonds and the superconducting transport mechanism
in them are unknown, therefore we cannot rule out a supercurrent between Al and superconducting states in \YRS{} other than the $A_{2u}$ helical phase.

\section{Ginzburg-Landau model of~superconductivity boosted~by~spin~density~wave}

\subsection{Magnetic and superconducting energy scales}

Electro-nuclear magnetic order develops in the AFM2 phase below $T_A = 1.5$\,mK.
Within the Landau model of AFM2~\cite{Knapp2023} the $T = 0$ free energy density is
\begin{align}
F_{\text{AFM2}} &= -A_{\mathrm{hf}} \mu_A / \sqrt{2} \cdot 0.18 \mu_N + \alpha \mu_A^2 + \gamma \mu_A^4
\notag\\ & = -3\,\mathrm{mJ/mol}.
\end{align}
Here $\mu_A = 0.09\mu_B$ is the amplitude of the electronic moments in the SDW.
After averaging over the SDW modulation and Yb isotopes the electronic moments $\mu_A / \sqrt{2}$
couple to nuclear moments $0.18\mu_N$ via the hyperfine constant $A_{\mathrm{hf}} = 102\,\mathrm{T}/\mu_B$.
The coefficients $\alpha = 0.026\,\mathrm{T}/\mu_B$, $\gamma = 0.2\,\mathrm{T}/\mu_B^3$ in the electronic internal energy have been obtained by fitting $T_A(H)$ to the mean-field model~\cite{Knapp2023}.

The Fermi liquid develops below $T_N = 70$\,mK with a large Sommerfeld coefficient, Eq.\ (1.15) in Ref.~\cite{LL:IX},
\begin{equation}
C_{\text{mol}} / T = V_{\text{mol}} \frac{m^* p_F}{3\hbar^3} k_B^2 = 1.65\,\mathrm{J/mol\,K^2}
\end{equation}
corresponding to the Fermi temperature $T_F \sim R / C_{\text{mol}} = 5$\,K.
Superconductivity develops below $T_c \sim 5$\,mK.
Since $T_c \ll T_F$, we use BCS formulae, regardless of the pairing mechanism.
For energy gap $\Delta = 10$\,mK the pairing energy, Eq.~(40.9) in \cite{LL:IX},
\begin{equation}
F_{\text{BCS}} = -V_{\text{mol}} \frac{m^* p_F}{4\pi^2 \hbar^3} \Delta^2 = -1\,\mu\mathrm{J/mol}
\end{equation}
In conclusion, $F_{\text{BCS}} \ll F_{\text{AFM2}}$, therefore
we can treat AFM2 order parameter as a temperature-dependent
external field in the Ginzburg-Landau theory of the superconductivity.

\onecolumngrid

\subsection{Ginzburg-Landau model}\label{sec:GL}

The model developed here illustrates how a triplet superconducting order parameter
$\vec d(\vec k)$ can react to an SDW. 
For simplicity we fix the momentum dependence $\tilde{\vec d}(\vec k)$
of the order parameter
\begin{equation}\label{eq:d:tilde}
\vec d(\vec k) = \Delta \tilde{\vec d}(\vec k)
\end{equation}
and only consider how SDW affects its amplitude $\Delta$.
As customary in the field of superfluid $^3$He,
here $\vec d(\vec k)$ directly describes the gap matrix~\cite{VWbook}.
We choose the scale of $\tilde{\vec d}(\vec k)$, such that $\Delta$
represents the maximum of the energy gap in the momentum space.
For the unitary order parameters~\cite{VWbook} we focus on,
this implies $\max\tilde{\vec d}^*(\vec k) \tilde{\vec d}(\vec k) = 1$.

In the absence of coupling to SDW, $\Delta$ is described by the conventional
Ginzburg-Landau free energy (per unit volume) for a scalar order parameter
\begin{equation}\label{eq:GL}
F_{\text{SC}} = \alpha(T - T_{c0}) |\Delta|^2 + \frac{u}{2} |\Delta|^4
\end{equation}
leading to the energy gap $|\Delta|^2 = {\alpha(T_{c0} - T)} / u$ at $T < T_{c0}$.
Here and below we do not include the Pauli or orbital suppression of the order parameter
by magnetic field in the model, and use the zero-field gap $\Delta$ as a measure
of the Pauli limiting field $H_P \propto \Delta$.
Our target is to qualitatively model the phase diagram up to in-plane fields
of order $H_P \sim 10$\,mT, which do not significantly affect the AFM2 SDW~\cite{Knapp2023}.
Therefore, when introducing this magnetic order parameter into the problem,
we again consider it at zero field.

\subsubsection{Competition between superconductivity and SDW}

Now we consider coupling between the superconducting order parameter $\vec d_0$
and the SDW of the AFM2 state
\begin{equation}
\vec\mSDW(\vec r) = \vec \mSDW_{\vec Q} \cos(\vec Q\cdot \vec r),
\end{equation}
where for convenience, we have taken the phase of the SDW to be zero.
AFM2 order is characterized by both nuclear and electronic staggered magnetizations.
Within the mean-field model~\cite{Knapp2023} these staggered magnetizations are collinear
and have the same $\vec Q$; close to $T_A$ their amplitudes have the same temperature dependence.
Therefore in the phenomenological model built here the same order parameter is used
to describe both staggered magnetizations,
ignoring any possible discrepancy in the temperature dependencies of their amplitudes far below $T_A$.

The leading-order coupling between $\vec d(\vec k)$ and the SDW
satisfying gauge and translational invariance is
\begin{equation}\label{eq:SC:SDW}
F_{\text{SC-SDW}} = \delta \mSDW_{\vec Q}^2 |\Delta|^2.
\end{equation}
Here $\delta$ may depend on the chosen superconducting state,
its orientation and the directions of $\vec\mSDW_{\vec Q}$ and $\vec Q$.
We remind the reader, that all these are fixed
in the model derived here, which only concerns with the temperature dependence of $\Delta(T)$.

The free energy
\begin{equation}\label{eq:GL:SDW}
F_{\text{SC}} + F_{\text{SC-SDW}} = \alpha\left(T - T_{c0} + \frac{\delta}{\alpha}\right)|\Delta|^2
+ \frac{u}{2}|\Delta|^4
\end{equation}
can be written in the simple form Eq.~\eqref{eq:GL} with effective critical temperature
$$
\widetilde T_{c0}(T) = T_{c0} - \frac{\delta}{\alpha} \mSDW_{\vec Q}(T)^2.
$$
Positive $\delta$ corresponds to a competition between the superconductivity and SDW, which
suppresses the superconductivity below $T_A$.
In case of $\delta < 0$ SDW boosts the superconductivity.



\subsubsection{PDW coupling between superconductivity and SDW}

We now consider the development of a spatially modulated component of the superconducting order parameter
\begin{equation}\label{eq:d0dq}
\vec d(\vec k, \vec r) = \vec d_0(\vec k) + \tfrac1 2 \vec d_{\vec q}(\vec k) e^{i\vec q \cdot \vec r}
+ \tfrac1 2 \vec d_{-\vec q}(\vec k) e^{-i\vec q \cdot \vec r}
\end{equation}
in response to the SDW.
Here $\vec d_0(\vec k)$ is the spatially uniform order discussed previously, while
$\vec d_{\pm \vec q}(\vec k)$ represent the PDW.
Microscopically these order parameters are related to the
pair amplitudes
\begin{align*}
\langle c_{-\vec k\alpha}c_{\vec k\beta}\rangle &\propto 
 i\big(\vec d_0(\vec k) \cdot \vec{\sigma}\sigma_2\big)_{\alpha \beta},\\
\langle c_{\vec q/2-\vec k\alpha}c_{\vec q/2+\vec k\beta}\rangle &\propto 
 i\big(\vec d_{\vec q}(\vec k) \cdot \vec{\sigma}\sigma_2\big)_{\alpha \beta},\\
\langle c_{-\vec q/2-\vec k\alpha}c_{-\vec q/2+\vec k\beta}\rangle &\propto 
 i\big(\vec d_{-\vec q}(\vec k) \cdot \vec{\sigma}\sigma_2\big)_{\alpha \beta}.
\end{align*}
Here $(\sigma_i)_{\alpha\beta}$ are the Pauli matrices and $c_{\vec k\alpha}$ is the 
annihilation operator for an electron with momentum $\vec k$ and spin orientation $\alpha$.

We describe the energy cost of sustaining the PDW by 
\begin{equation}\label{eq:kinetic}
F_{\text{PDW}} = \tfrac 1 2 \chi\langle\vec d_{\vec q}^* \cdot \vec d_{\vec q}^{\vphantom*} + 
\vec d_{-\vec q}^* \cdot \vec d_{-\vec q}^{\vphantom*}\rangle\FS.
\end{equation}
If $q \gg 1/\xi$, $F_{\text{PDW}}$ is dominated by the kinetic energy of the PDW. For instance taking
$$F_{\nabla} = K \big\langle \big(\partial d^*_{\alpha}(\vec k,\vec r)/\partial r_i \big)
\big(\partial d^{\vphantom*}_{\alpha}(\vec k,\vec r)/\partial r_i \big)\big\rangle_{\vec k,\vec r}
$$
leads to $\chi = Kq^2/2$.
In general the Ginzburg-Landau theory of spin-triplet order parameters includes
multiple gradient terms~\cite{VWbook}, thus $\chi$ may depend on the orientation
of $\vec q$ for a given $\vec d_{\vec q}(\vec k)$.
A system with a marked tendency towards PDW formation will have a small PDW susceptibility $\chi$.

The modulated spin-triplet superconductor is characterized by the spin density 
\begin{align}
\vec\mSC(\vec r) & \propto i\langle \vec d^*(\vec k, \vec r) \times
\vec d(\vec k, \vec r)\rangle\FS \notag\\
&= i\big\langle \vec d_0^* \times \vec d_0^{\vphantom*} 
    + \tfrac 1 4\vec d_{\vec q}^* \times \vec d_{\vec q}^{\vphantom*}
    + \tfrac 1 4\vec d^*_{-\vec q} \times \vec d_{-\vec q}^{\vphantom*}\big\rangle\FS 
    + \tfrac 1 2 ie^{i \vec q \cdot \vec r}\left\langle \vec d_0^* \times \vec d_{\vec q}^{\vphantom*}
    + \vec d_{-\vec q}^* \times \vec d_0^{\vphantom*}\right\rangle\FS \notag\\
&~~~+ \tfrac 1 2 ie^{-i \vec q \cdot \vec r}\left\langle \vec d_0^* \times \vec d_{-\vec q}^{\vphantom*}
    + \vec d_{\vec q}^* \times \vec d_0^{\vphantom*}\right\rangle\FS 
    + \tfrac 1 4 ie^{2i\vec q\cdot\vec r} \left\langle\vec d_{-\vec q}^*
		\times\vec d_{\vec q}^{\vphantom*}\right\rangle\FS
    + \tfrac 1 4 ie^{-2i\vec q \cdot\vec r}\left\langle\vec d_{\vec q}^*
		\times\vec d_{-\vec q}^{\vphantom*}\right\rangle\FS.
\end{align}
We consider the leading order coupling between $\vec\mSDW$ and $\vec\mSC$
\begin{equation}\label{eq:fPDW0}
F_{\text{SDW-SC-PDW}} \propto \langle \vec\mSDW(\vec r) \cdot \vec\mSC(\vec r) \rangle_{\vec r}.
\end{equation}
In the case of a superconducting condensate heavily hybridized with Yb f-electrons, 
a possible microscopic mechanism leading to this interaction is the hyperfine coupling to the nuclear SDW.

We are interested in the coupling between $\vec d_0$ and $\vec\mSDW_{\vec Q}$ mediated by $\vec d_{\pm q}$.
Then for the spatial average $\langle \dots \rangle_{\vec r}$ in Eq.~\eqref{eq:fPDW0} to be non-zero,
the PDW must have the same wavevector as the SDW
(for both modulated orders we assume long wavelengths in comparison to the interatomic spacing).
Substituting $\vec q = \vec Q$ in Eq.~\eqref{eq:fPDW0} we get
\begin{equation}\label{eq:SDW:SC:PDW}
F_{\text{SDW-SC-PDW}} = \tfrac 1 2 \lambda \vec \mSDW_{\vec Q} \cdot
	\big\langle \vec d_0^* \times (\vec d_{-\vec Q}^{\vphantom*} + \vec d_{\vec Q}^{\vphantom*})
	+ (\vec d_{-\vec Q}^* + \vec d_{\vec Q}^{*}) \times \vec d_0^{\vphantom*} \big\rangle\FS.
\end{equation}
Using tensor notation and replacing $\langle \dots \rangle\FS$ with a summation over
momentum states we rewrite Eq.~\eqref{eq:SDW:SC:PDW} as
\begin{align}
F_{\text{PDW}} &+ F_{\text{SDW-SC-PDW}} = \frac{1}{\sum\limits_{\vec k} 1}
\sum_{\vec k} \Big\{ \tfrac 1 2 \chi\big[ d_{\vec  Q}^{*\gamma} d_{\vec  Q}^\gamma
                                        + d_{-\vec Q}^{*\gamma} d_{-\vec Q}^\gamma\big] \notag\\
               & + \tfrac 1 2 i\lambda \big[\varepsilon_{\alpha\beta\gamma} \mSDW_{\vec Q}^\alpha
                                            d_0^{*\beta} (d_{\vec Q}^\gamma + d_{-\vec Q}^\gamma)
                                         - \varepsilon_{\alpha\beta\gamma} \mSDW_{\vec Q}^\alpha
                                            d_0^{\beta} (d_{\vec Q}^{*\gamma} + d_{-\vec Q}^{*\gamma})
                                           \big]\Big\}.
\end{align}
By minimising $F_{\text{PDW}} + F_{\text{SDW-SC-PDW}}$ with respect to $d_{\pm\vec Q}^{*\gamma}$
we find the PDW order parameter
\begin{equation}\label{eq:dQ}
\vec d_{\vec Q}(\vec k) = \vec d_{-\vec Q}(\vec k)
  = \frac{i\lambda}{\chi} \vec \mSDW_{\vec Q} \times \vec d_0(\vec k),
\end{equation}
so Eq.~\eqref{eq:d0dq} can be written as
\begin{equation}
\vec d(\vec k, \vec r) = \vec d_0(\vec k) + \vec d_{\vec Q}(\vec k) \cos(\vec Q\cdot\vec r).
\end{equation}
Thus we find the PDW to be of Larkin-Ovchinnikov rather than Fulde-Ferrell type.
The PDW modulation is in phase with SDW, while the phase of $\vec d_{\vec Q}(\vec k)$ is
shifted by $\pi/2$ relative to $\vec d_0(\vec k)$.
Substituting Eq.~\eqref{eq:dQ} into Eqs.~\eqref{eq:kinetic} and \eqref{eq:SDW:SC:PDW}
we find the energy gain from the formation of PDW
\begin{equation}\label{eq:fPDW}
F_{\text{PDW}} + F_{\text{SDW-SC-PDW}} = -\frac{\lambda^2}{\chi} \big\langle
|\vec\mSDW_{\vec Q} \times \vec d_0|^2 \big\rangle\FS \le 0.
\end{equation}
Finally, expressing $\vec d_0$ in the form Eq.~\eqref{eq:d:tilde}, we rewrite Eq.~\eqref{eq:fPDW} as
\begin{equation}\label{eq:fPDWtilde}
F_{\text{PDW}} + F_{\text{SDW-SC-PDW}} = -\frac{\lambda^2 \big\langle | \hvec\unitmSDW_{\vec Q} \times
\tilde{\vec d}|^2\big\rangle_{\vec k}}{\chi} \mSDW_{\vec Q}^2 |\Delta|^2
= -\tilde{\delta} M_{\vec Q}^2 |\Delta|^2,
\end{equation}
where 
\begin{equation}
\tilde \delta =  \frac{\lambda^2 \big\langle | \hvec\unitmSDW_{\vec Q} \times
\tilde{\vec d}|^2\big\rangle_{\vec k}}{\chi} \ge 0
\end{equation}
defines the magnitude of the attraction induced by PDW fluctuations, and 
$\hvec\unitmSDW_{\vec Q}$ is the unit vector in the direction of $ {\vec\mSDW}_{\vec Q}$.
This attractive coupling will only vanish
if $\vec d_0 \parallel \vec\mSDW_{\vec Q}$ at all $\vec k$.
In particular $\tilde\delta = 0$ for the Pauli-unlimited nematic phase
($\hvec d \perp \vec H, \hvec d \perp c$) .


The combined terms of our Ginzburg-Landau model are then
\begin{align}
F &= F_{\text{SC}} + F_{\text{SDW-SC}} + F_{\text{PDW}} + F_{\text{SDW-SC-PDW}}
  = \alpha\left(T - T_{c0} - \frac{\tilde\delta -\delta}{\alpha} \mSDW_{\vec Q}^2(T)\right)|\Delta|^2
  + \frac{u}{2}|\Delta|^4.
\end{align}
We see that provided $\tilde \delta > \delta$, a condition satisfied for sufficiently soft PDW fluctuations, the effective $T_c$ is boosted according to 
\begin{equation}
T_c = T_{c0}+\frac{\tilde\delta - \delta}{\alpha} \mSDW_{\vec Q}^2(T)
\end{equation}
By minimizing the free energy we get the corresponding energy gap
\begin{equation}
|\Delta(T)|^2 = \frac{\alpha}{u}\left(T_{c0} +
	\frac{\tilde \delta - \delta}{\alpha} \mSDW_{\vec Q}^2(T) - T\right).
\end{equation}


\subsection{Details of the model shown in Figure 4}

In this section all quantities are dimensionless,
as the model is aimed at reproducing qualitative features of the experimental phase diagrams.
We take a simple ansatz for AFM2 order parameter
$$
\mSDW_{\vec Q} = \tanh^{0.07} (1 - T/T_A) \qquad \text{at $T < T_A$},
$$
that reproduces the saturation below $0.5T_A$ and critical exponent
exhibited by the staggered electronic magnetization~\cite{Knapp2023}.
The temperature scale is set by taking $T_A = 1$.

To introduce the sample nonuniformity in the model,
we allow the intrinsic (i.e. in the absence of AFM2 SDW) critical temperature $T_{c0}$ to vary.
We take Ginzburg-Landau coefficients $\alpha = 1/T_{c0}$, $u = 1/T_{c0}^2$,
these dependencies on $T_{c0}$ are found in the Ginzburg-Landau theory of $^3$He,
when the order parameter is expressed in the units of energy gap~\cite{VWbook}.

The competition with SDW is modeled by $\delta = 2$ for easy-plane helical and nematic states.
The boost via PDW is $\tilde\delta = 3$ for the helical and $\tilde\delta = 0$ for the nematic state.
For the helical phase the amplitude of PDW (in arbitrary units) is computed as
$\Delta_{\vec Q} = \mSDW_{\vec Q} \Delta$.

\subsubsection{Alternative scenario with easy-axis states}

In the alternative scenario an easy-axis state, such as the chiral phase, replaces
the easy-plane nematic state. In this case the boost via Eq.~\eqref{eq:fPDWtilde}
is present, in contrast to the easy-plane nematic state.
We require the model to account for the observed re-entrant normal state at $T < T_A$,
so the easy-axis state must be suppressed below $T_A$ via Eq.~\eqref{eq:GL:SDW}
with a sufficiently large positive $\delta$.

In the case of the chiral phase with $\tilde{\vec d}(\vec k) = \hvec c(k_a \pm ik_b)$,
the boost is twice as strong as for the helical phase, corresponding to $\tilde\delta = 6$.
Thus in order to reproduce Fig.~4D, where $\delta - \tilde\delta = 2$,
for the chiral state we must take $\delta = 8$,
four times stronger than the value we adopt for the helical phase.

\end{document}